\documentclass[12pt]{article}
\usepackage{amssymb,amsmath}
\usepackage{comment}
\usepackage{bm}
\usepackage{enumerate}
\usepackage{graphicx}
\usepackage{hyperref}

\usepackage{enumitem}

\usepackage{soul}
\usepackage{cancel}
\usepackage[normalem]{ulem}

\newcommand{\Ncal}{\mathcal{N}}
\newcommand{\del}{\partial}

\newcommand{\psib}{\bar{\psi}}
\newcommand{\thetab}{\bar{\theta}}
\newcommand{\phib}{\bar{\phi}}

\newcommand{\xib}{\bar{\xi}}
\newcommand{\Qb}{\overline{Q}}
\newcommand{\Db}{\overline{D}}
\newcommand{\deldel}[2]{\frac{\del #1}{\del #2}}

\usepackage[usenames,dvipsnames,svgnames,table]{xcolor}
\definecolor{darkblue}{cmyk}{0.9,0.9,0,0}
\hypersetup{colorlinks=false, linkcolor=darkblue,linkcolor=darkblue, citecolor=darkblue}

\numberwithin{equation}{section}

\begin{document}

\thispagestyle{empty}
\begin{flushright}
KIAS-P16058\\
NCTS-TH/1605


\end{flushright}
\vskip2cm
\begin{center}
{\Large $(2,2)$ and $(0,4)$ Supersymmetric Boundary Conditions \\in 3d $\mathcal{N}=4$ Theories and Type IIB Branes}

\vskip1.5cm
Hee-Joong Chung\footnote{hjchung@kias.re.kr}$^{a}$ and Tadashi Okazaki\footnote{tadashiokazaki@phys.ntu.edu.tw}$^{b}$

\bigskip
{\it 
${}^a$Korea Institute for Advanced Study,
Seoul 02455, Republic of Korea

${}^b$Department of Physics and Center for Theoretical Sciences,\\
National Taiwan University, Taipei 10617, Taiwan}
\\

\end{center}

\vskip1cm
\begin{abstract}
The half-BPS boundary conditions preserving $\mathcal{N}=(2,2)$ and $\mathcal{N}=(0,4)$ supersymmetry in 3d $\mathcal{N}=4$ supersymmetric gauge theories are examined. 
The BPS equations admit decomposition of the bulk supermultiplets into specific boundary supermultiplets of preserved supersymmetry. 
Nahm-like equations arise in the vector multiplet BPS boundary condition preserving $\mathcal{N}=(0,4)$ supersymmetry
and Robin-type boundary conditions appear for the hypermultiplet coupled to vector multiplet when $\mathcal{N}=(2,2)$ supersymmetry is preserved. 
The half-BPS boundary conditions are realized in the brane configurations of Type IIB string theory. 
\end{abstract}


\newpage
\setcounter{tocdepth}{2}
\tableofcontents

\section{Introduction}
\label{0sec}
The boundary conditions for supersymmetric field theory preserving a part of supersymmetries of the original bulk theory provide important new ingredients and insights to the original system, 
for example, the description of branes in string/M-theory and in target space of the field theories, dualities or holography in the presence boundary conditions, mirror symmetry, and also Geometric Langlands Program.
The supersymmetric (SUSY) boundary conditions have been studied in a number of contexts such as 
2d $\mathcal{N}=(2,2)$ theories, 
\cite{Ooguri:1996ck,Hori:2000ck,Herbst:2008jq}
4d $\mathcal{N}=4$ theories, 
\cite{Gaiotto:2008sa,Gaiotto:2008ak,Hashimoto:2014nwa,Hashimoto:2014vpa}
3d $\mathcal{N}=2$ theories, 
\cite{Okazaki:2013kaa,Aprile:2016gvn}
and 3d $\mathcal{N}=4$ theories, 
\cite{Bullimore:2016nji}
and BLG and ABJM theories 
\cite{Berman:2009xd,Hosomichi:2014rqa,Okazaki:2015fiq}.

In this paper, we study the half-BPS boundary conditions of 3d $\mathcal{N}=4$ supersymmetric theories 
preserving $\mathcal{N}=(2,2)$ and $\mathcal{N}=(0,4)$ supersymmetry at the boundary, 
which we call A-type and B-type, respectively. 
We explicitly calculate the boundary BPS equations 
for the 3d $\mathcal{N}=4$ pure vector multiplet, pure hypermultiplet, 
hypermultiplet coupled to vector multiplet such as SQCD, 
and also its supersymmetric deformations by Fayet-Iliopoulos (FI) parameters and mass parameters. 
For each A- and B-type boundary conditions for the vector multiplet
we have two sets of boundary conditions, 
which we call ``electric-like" and ``magnetic-like". 
Interestingly, the half-BPS boundary conditions preserving $\mathcal{N}=(0,4)$ for the vector multiplet include Nahm-like equation.
For the hypermultiplet coupled to vector multiplet, 
we see that a certain types of Robin boundary conditions arise. 
By studying the BPS equations, we read off the boundary degrees of freedom arising from the bulk 3d $\mathcal{N}=4$ vector multiplet and hypermultiplet. 
These are discussed in section \ref{sec3dn4a}.

In section \ref{secbrane}, 
we propose the brane configurations corresponding to the boundary conditions of 3d $\mathcal{N}=4$ theories preserving $\mathcal{N}=(2,2)$ and $\mathcal{N}=(0,4)$ supersymmetry 
by introducing additional branes to the brane configuration of Hanany and Witten in Type IIB string theory \cite{Hanany:1996ie} realizing 3d $\mathcal{N}=4$ theories. 
We give a remark on the map of the boundary degrees of freedom from the bulk supermultiplet under $S$-duality of Type IIB string theory.

In section \ref{seccon}, we summarize our results and discuss future directions.

\section{Half-BPS boundary conditions in 3d $\mathcal{N}=4$ theories}
\label{sec3dn4a}

In this section, we consider the $(2,2)$ or $(0,4)$-preserving boundary conditions for pure vector multiplet, pure hypermultiplets, and hypermultiplets coupled to vector multiplet with FI and mass deformations.
We also see the decomposition of the bulk supermultiplet at the boundary as supermultiplets of preserved supersymmetries.

\subsection{Vector multiplet}
\label{secvm1}
In this subsection we study the half-BPS boundary conditions for the 3d $\mathcal{N}=4$ vector multiplet. 
The 3d $\mathcal{N}=4$ vector multiplet contains 
a three-dimensional gauge field $A_{\mu}$, $\mu=0,1,2$, 
three real scalar fields $\phi^{i}$, $i=3,4,5$,
an auxiliary field $F$, 
and a fermionic field $\lambda$. 
They are in the adjoint representation of the gauge group $G$ 
and transform as 
\begin{align}
\begin{split}
\label{3dn4vm1a}
A_{\mu}&: (\bm{3},\bm{1},\bm{1})	\\
\phi^{i}&: (\bm{1},\bm{3},\bm{1})	\\
F&: (\bm{1},\bm{1},\bm{3})	\\
\lambda&: (\bm{2},\bm{2},\bm{2})
\end{split}
\end{align}
under the $SO(1,2)\times SO(3)_{C}\times SO(3)_{H}$. 
The 3d $\mathcal{N}=4$ supersymmetric field theories have the R-symmetry group 
$SO(4)_{R}\cong SU(2)_{C}\times SU(2)_{H}$ 
where the $SU(2)_{C}$ (resp. $SU(2)_{H}$) is the double cover of the $SO(3)_{C}$ (resp. $SO(3)_{H}$).

The 3d $\mathcal{N}=4$ vector multiplet can be expressed as 3d $\mathcal{N}=2$ vector multiplet $V(A_{\mu},\sigma,\lambda,D)$ and adjoint chiral multiplet $\Phi(\phi,\psi_{\phi},F_\phi)$.
Our notations for the 3d $\mathcal{N}=2$ superspace and supermultiplet are summarized in the Appendix \ref{appss}.
The action of the 3d $\mathcal{N}=4$ vector multiplet in terms of 3d $\mathcal{N}=2$ supermultiplets is given by
\begin{align}
\label{3dn4vm1c}
S^{\mathcal{N}=4}_{V}&=S_{V}^{\mathcal{N}=2}+S_{\Phi}^{\mathcal{N}=2}
\end{align}
with 
\begin{align}
\label{3dn4vm1d}
S_{V}^{\mathcal{N}=2}&=\frac{1}{g_{\text{3d}}^{2}}
\int d^{3}x d^{4}\theta\ \mathrm{Tr} (\Sigma^{2}),
\\
\label{3dn4vm1e}
S_{\Phi}^{\mathcal{N}=2}&=-\frac{1}{g_{\text{3d}}^{2}}\int d^{3}x d^{4}\theta\ \mathrm{Tr} (\overline{\Phi} e^{-2V}\Phi e^{2V}).
\end{align}
where $\Sigma$ is a linear multiplet.
In component, they are
\begin{align}
\begin{split}
S_{V}^{\mathcal{N}=2} = \frac{1}{g_{\text{3d}}^2} \int d^3 x \ \text{Tr} 
\Big[ 
&-\frac{1}{4} F_{\mu \nu} F^{\mu \nu} F^{\mu \nu} -\frac{1}{2} D^\mu \sigma D_\mu \sigma + \frac{1}{2} D^2
- i \overline{\lambda} \sigma^\mu D_\mu \lambda + i \lambda [\sigma, \overline{\lambda}]
\Big]	\,	,
\end{split}	\\
\begin{split}
S_{\Phi}^{\mathcal{N}=2} = \frac{1}{g_{\text{3d}}^2} \int d^3 x \ \text{Tr} 
\Big[
&-D_\mu \overline{\phi} D^\mu \phi - i \overline{\psi} \sigma^\mu D_\mu \psi + \overline{F} _\phi F_\phi		\\
&+ \overline{\phi} [\phi, D] - \sqrt{2} i \psi [\overline{\phi}, \lambda] + \sqrt{2} i \overline{\psi} [\phi, \overline{\lambda}]
+ i \overline{\psi} [\psi, \sigma] - \overline{\phi} [\sigma, [\sigma, \phi]]		
\Big]	\,	,
\end{split}
\end{align}
respectively.
The actions are invariant under the supersymmetry transformations
\begin{align}
\label{3dn2vmsusy1}
\delta A_{\mu}&=i\overline{\xi}\sigma_{\mu}\lambda
+i\xi \sigma_{\mu}\overline{\lambda},\\
\label{3dn2vmsusy2}
\delta \sigma&=\xi \overline{\lambda}
-\overline{\xi}\lambda,\\
\label{3dn2vmsusy3}
\delta \lambda&=i\xi D
-\frac12 \gamma^{\mu\nu}\xi F_{\mu\nu}
-i\gamma^{\mu}\xi D_{\mu}\sigma,\\
\label{3dn2vmsusy4}
\delta \overline{\lambda}
&=-i\overline{\xi}D-\frac12 \gamma^{\mu\nu}\overline{\xi}F_{\mu\nu}
+i\gamma^{\mu}\overline{\xi}D_{\mu}\sigma,\\
\label{3dn2vmsusy5}
\delta D&= -\xi \sigma^{\mu}D_{\mu}\overline{\lambda}
+\overline{\xi}\sigma^{\mu}D_{\mu}\lambda+
\xi[\sigma,\overline{\lambda}]+
\overline{\xi}[\sigma,\lambda]
\end{align}
for 3d $\mathcal{N}=2$ vector multiplet $V$ with the Wess-Zumino gauge, and 
\begin{align}
\label{3dn2cmadj1}
\delta \phi&=\sqrt{2}\xi \psi_{\phi},\\
\label{3dn2cmadj2}
\delta \psi_{\phi}&=\sqrt{2}i\gamma^{\mu}\overline{\xi} D_{\mu}\phi
+\sqrt{2}\xi F-\sqrt{2}i\overline{\xi}[\sigma,\phi],\\
\label{3dn2cmadj3}
\delta F&=
\sqrt{2}i\overline{\xi}\sigma^{\mu}D_{\mu}\psi_{\phi}
+2i\overline{\xi}[\overline{\lambda},\phi]
-\sqrt{2}i\overline{\xi}[\psi_{\phi},\sigma]	\,	,
\end{align}
for the 3d $\mathcal{N}=2$ adjoint chiral multiplets $\Phi$.

Suppose we have a boundary in $x^{2}$ direction, say at $x^{2}=0$. 
Employing the Noether method, 
we find the normal component $J^{2}$ of supercurrent 
of the 3d $\mathcal{N}=4$ vector multiplet 
in terms of 3d $\mathcal{N}=2$ language
from the action \eqref{3dn4vm1c} 
and the supersymmetric transformations \eqref{3dn2vmsusy1}-\eqref{3dn2cmadj3}, 
\begin{align}
\begin{split}
J^{2} 
&= J^{2}_{\text{vec}} + J^{2}_{\text{adj}}	\\
&= -\frac{1}{4} i \xi^{2 {m}{n}} F_{{m}{n}} \lambda
+ \frac{1}{2} i F^{{m} 2}  \sigma_{m} \lambda
-\frac{1}{2} D_{m} \sigma \sigma^{{m} 2} \lambda  
+ \frac{1}{2} D^2 \sigma \lambda  	
\\
& \quad +\frac{1}{\sqrt{2}} D^2 \phi \overline{\psi}_{\phi} 
- \frac{1}{\sqrt{2}} D_{m} \phi \sigma^{{m} 2} \overline{\psi}_{\phi} 
+ \frac{1}{2} [\overline{\phi}, \phi] \sigma^2 \lambda 
- \frac{1}{\sqrt{2}} [\sigma, \phi] \sigma^2 \overline{\psi}_{\phi}		
\end{split}	\label{susycurrent}	
\end{align}
in the on-shell
where $m,n,\cdots=0,1$ are space-time indices of two-dimensional boundary.\footnote{
We can put the supercurrent of 3d $\mathcal{N}=2$ vector and adjoint chiral multiplet in $SU(2)_C \times SU(2)_H$ manifest expression, which leads to the 3d $\mathcal{N}=4$ manifest supercurrent for the 3d $\mathcal{N}=4$ vector multiplet. Denoting 3d $\mathcal{N}=4$ fermions and scalars by 
\begin{align}
\lambda^{\alpha A \dot{A}} = \left( \begin{matrix} \lambda^{\alpha} & -\psi^{\alpha}_\phi \\ \overline{\psi}^{\alpha}_\phi & \overline{\lambda}^{\alpha} \end{matrix} \right)	\,	,	\quad
\phi^{A}_{\ B} = \left( \begin{matrix} \sigma & \sqrt{2}\phi \\ \sqrt{2}\overline{\phi} & -\sigma \end{matrix} \right)	\,	,
\end{align}
respectively, the normal component of the supercurrent can be written as
\begin{align}
\begin{split}
J^2_{\text{vec}} + J^2_{\text{adj}} 
 = J^2 \ &= \  (J^{2})^{\alpha A \dot{A}}	\\
&= \
\frac{1}{2} i F_{01}  \lambda^{\alpha A \dot{A}} 
-\frac{1}{2} D^2 \phi^{A}_{\ B}  \lambda^{\alpha B \dot{A}} 
+\frac{1}{2} i F^{2}_{\ m}  (\gamma^{m})^{\alpha}_{\ \beta} \lambda^{\beta A \dot{A}}		\\
& \quad +\frac{1}{4} [\phi^{A}_{\ C}, \phi^{C}_{\ B}]  (\gamma^{m})^{\alpha}_{\ \beta} \lambda^{\beta B \dot{A}}
+\frac{1}{2} D_{m} \phi^{A}_{\ B}  (\gamma^{m2})^{\alpha}_{\ \beta} \lambda^{\beta B \dot{A}}	\,	.
\end{split}
\end{align}
Here, $A$ and $\dot{A}$ are indices for $SU(2)_C$ and $SU(2)_H$, respectively.
}

In the presence of boundary, the translation invariance is broken, so the supersymmetry is broken in general.
However, some of supersymmetry can be preserved at the boundary by imposing specific boundary conditions, \textit{i.e.} the supersymmetric or BPS boundary conditions. 
The BPS boundary conditions can be found by demanding that the normal component of the supercurrent at the boundary vanishes
\begin{align}
\label{susybc1a}
\begin{split}
0
=& \ \overline{\xi}J^{2}-\xi \overline{J}^{2}	\\
=& \ -\frac{1}{4} i \xi^{2 {m}{n}} F_{{m}{n}} (\overline{\xi} \lambda) 
+ \frac{1}{2} i F^{{m} 2} (\overline{\xi} \sigma_{m} \lambda)
-\frac{1}{2} D_{m} \sigma (\overline{\xi} \sigma^{{m} 2} \lambda) 
+ \frac{1}{2} D^2 \sigma (\overline{\xi} \lambda) 	
\\
&-\frac{1}{4} i \xi^{2 {m}{n}} F_{{m}{n}} (\xi \overline{\lambda}) 
+ \frac{1}{2} i F^{{m} 2} (\xi \sigma_{m} \overline{\lambda})
+\frac{1}{2} D_{m} \sigma (\xi \sigma^{{m} 2} \overline{\lambda}) 
- \frac{1}{2} D^2 \sigma (\xi \overline{\lambda}) 	
\\
&
+\frac{1}{\sqrt{2}} D^2 \phi (\overline{\xi} \overline{\psi}_{\phi}) 
- \frac{1}{\sqrt{2}} D_{m} \phi (\overline{\xi} \sigma^{{m} 2} \overline{\psi}_{\phi})
+ \frac{1}{2} [\overline{\phi}, \phi] (\overline{\xi} \sigma^2 \lambda) 
- \frac{1}{\sqrt{2}} [\sigma, \phi] (\overline{\xi} \sigma^2 \overline{\psi}_{\phi})		\\
&- \frac{1}{\sqrt{2}} D^2 \overline{\phi} (\xi \psi_{\phi}) 
+\frac{1}{\sqrt{2}} D_{m} \overline{\phi} (\xi \sigma^{{m} 2} \psi_{\phi}) 
- \frac{1}{2} [\overline{\phi}, \phi] (\xi \sigma^2 \overline{\lambda})
- \frac{1}{\sqrt{2}} [\sigma, \overline{\phi}] (\xi \sigma^2 \psi_{\phi}) 	\, .	
\end{split}	
\end{align}
where we impose the boundary condition on fermions such that the bulk equations of motion are still satisfied.

Although there are various solutions to the supersymmetric boundary conditions (\ref{susybc1a}), in this paper we will focus on the half-BPS boundary conditions preserving $\mathcal{N}=(2,2)$ and $\mathcal{N}=(0,4)$ supersymmetry at the boundary, 
which we call A- and B-type boundary conditions, respectively \cite{Ooguri:1996ck,Okazaki:2013kaa}.

\subsubsection{A-type boundary conditions}
\label{secvma}

For the A-type boundary conditions, the supersymmetric parameter $\xi$ satisfies the projection condition\footnote{
Since the projection condition \eqref{asusybc1} is written in terms of 3d $\mathcal{N}=2$ SUSY parameters, it leads to $\mathcal{N}=(1,1)$ SUSY parameters at the boundary. But with the supersymmetry enhancement to 3d $\mathcal{N}=4$ in mind, as far as bosonic boundary BPS equations are concerned, it is okay to work with \eqref{asusybc1} for convenience.
}
\begin{align}
\gamma^2 \xi = \overline{\xi} \, .
\label{asusybc1}
\end{align}
To find the bosonic boundary conditions from (\ref{susybc1a}), we choose the fermionic boundary conditions
\begin{align}
\gamma^2 \lambda = e^{2i\theta} \overline{\lambda} \,		\qquad		\gamma^2 \psi_{\phi} = e^{2i\theta} \overline{\psi}_{\phi}	\, .
\label{asusybc2}
\end{align}
where $\theta\in \mathbb{R}$ is a constant parameter. 
Note that this form of fermionic boundary conditions is compatible with the bulk equations of motion for fermions $\lambda$ 
and $\psi_\phi$. 
From (\ref{asusybc1}) and (\ref{asusybc2}), we obtain 
\begin{align}
\begin{split}
\overline{\xi}\lambda&= -e^{2i\theta}\xi \overline{\lambda}	\,	, 
\hspace{30mm} \overline{\xi} \sigma_{m} \lambda= e^{2i\theta}\xi \sigma_{m} \overline{\lambda}	\,	, 	\\
\overline{\xi} \sigma^{{m} 2} \lambda&= e^{2i\theta} \xi \sigma^{{m} 2} \overline{\lambda}	\,	, 
\hspace{27.5mm} \overline{\xi} \sigma^2 \lambda= - e^{2i\theta} \xi \sigma^2 \overline{\lambda}	\,	; 	
\label{a1r1}	
\end{split}	\\
\begin{split}
\overline{\xi} \overline{\psi}_{\phi}&= -e^{-2i\theta}\xi \psi_{\phi}	\,	, 
\hspace{23mm} \overline{\xi} \sigma_{m} \overline{\psi}_{\phi}= e^{-2i\theta}\xi \sigma_{m} \psi_{\phi}	\,	, 	\\
\overline{\xi} \sigma^{{m} 2} \overline{\psi}_{\phi}&= e^{-2i\theta} \xi \sigma^{{m} 2} \psi_{\phi}	\,	, 
\hspace{20.5mm}\overline{\xi} \sigma^2 \overline{\psi}_{\phi} =- e^{-2i\theta} \xi \sigma^2 \psi_{\phi}
\label{a1r2}
\end{split}
\end{align}
With the above fermionic boundary conditions and the above formulae (\ref{a1r1})-(\ref{a1r2}), one can rewrite the general supersymmetric boundary conditions (\ref{susybc1a}) as
\begin{align}
\begin{split}
0&=
\frac{1}{4}
 i \xi^{2mn} F_{mn} (e^{2i\theta} - 1) (\xi \overline{\lambda}) 
- \frac{1}{2} i F^{2m} (e^{2i\theta} + 1) (\xi \sigma_{m} \overline{\lambda}) 	\\
&-\frac{1}{2} D_{m} \sigma (e^{2i \theta} -1) (\xi \sigma^{m 2} \overline{\lambda})  
-\frac{1}{2} D^2 \sigma (e^{2i\theta}+1) (\xi \overline{\lambda}) 
\\
&
-\frac{1}{\sqrt{2}} (e^{-2i\theta} D^2 \phi + D^2 \overline{\phi}) (\xi \psi_{\phi}) - \frac{1}{\sqrt{2}} (e^{-2i \theta} D_{m} \phi - D_{m} \overline{\phi}) (\xi \sigma^{m 2} \psi_{\phi})		\\
&
- \frac{1}{2} [\overline{\phi}, \phi] (e^{2 i \theta} +1) (\xi \sigma^2 \overline{\lambda}) 
+ \frac{1}{\sqrt{2}} ( [\sigma, \phi] e^{-2i\theta} - [\sigma, \overline{\phi}] ) (\xi \sigma^2 \psi_{\phi})	\, .
\end{split}	\label{asusybc3}
\end{align}

Without further projection conditions, we can find supersymmetric bosonic configurations 
as the non-trivial solutions to (\ref{asusybc3}) when $\theta=0$ and $\frac{\pi}{2}$.

From now on, we often identify the scalars $\sigma$ and $\phi, \overline{\phi}$ of 3d $\mathcal{N}=2$ vector and adjoint chiral multiplet with the scalars $\phi^i$, $i=3,4,5$ of the 3d $\mathcal{N}=4$ vector multiplet as
\begin{align}
\sigma = \phi^3			\,	,	\quad
\text{Re} \ \phi = \phi^4	\,	,	\quad
\text{Im} \ \phi = \phi^5	\,	.	\label{iden}
\end{align}

\begin{itemize}[leftmargin=5mm]
\item[i)] $\gamma^2 \lambda = \overline{\lambda}$ and $\gamma^2 \psi_\phi = \overline{\psi}_\phi$ ($\theta = 0$)

From \eqref{asusybc3} with $\theta=0$, we find the boundary conditions
\begin{align}
\label{asusybc3c}
F_{2m}&=0	\,	,\\
\label{asusybc3d}
D_{2}\phi^{a}&=0	\,	,\\
\label{asusybc3d0}
D_{m}\phi^{5}&=0	\,	,\\
\label{asusybc3e}
[\phi^{a},\phi^{5}]&=0	\,	,
\end{align}
where $a = 3, 4$. 
The two-dimensional gauge field $A_{m}$ and the two scalar fields $\phi^{a}$ satisfy Neumann-like boundary conditions \eqref{asusybc3c} and \eqref{asusybc3d}, so can fluctuate at the boundary. 
The condition \eqref{asusybc3c} can be thought as the Dirichlet-like condition for the scalar field $A_{2}$. 
The scalar field $\phi^{5}$ satisfy the Dirichlet-like condition \eqref{asusybc3d0}.
In particular, (\ref{asusybc3d0}) and (\ref{asusybc3e}) can be solved by setting $\phi^{5}=0$. 
We call the above set of boundary conditions (\ref{asusybc3c})-(\ref{asusybc3e}) the \textit{electric-like} A-type boundary conditions where the electric-like field $E_{m}=F_{2m}$ generated by scalar potential $A_{2}$ is required to be constant while the magnetic-like field $B=F_{01}$ can fluctuate at the boundary.

\item[ii)] $\gamma^2 \lambda = -\overline{\lambda}$ and $\gamma^2 \psi_\phi = -\overline{\psi}_\phi$ ($\theta= \frac{\pi}{2}$)

In this case, the boundary conditions read
\begin{align}
\label{asusybc3a1}
F_{01}&=0	\,	,\\
\label{asusybc3a2}
D_{2}\phi^{5}&=0	\,	,\\
\label{asusybc3b}
D_{m}\phi^{a}&=0	\,	,\\
\label{asusybc3b0}
[\phi^{a},\phi^{b}]&=0	\,	.
\end{align}
where $a, b, \cdots = 3, 4$.
We obtain the Dirichlet-like boundary condition for the two-dimensional gauge field $A_{m}$ and the Neumann-like boundary condition for the scalar field $\phi^{5}$.
The third equation (\ref{asusybc3b}) is the Dirichlet-like condition on two scalar fields $\phi^{a}$. 
The last constraint (\ref{asusybc3b0}) implies that the two scalar fields satisfying Dirichlet-like condition commute with each other. 
So one possible solution is to set them zero at the boundary. 
Meanwhile, as there is no constraint on $F_{m 2}$, the scalar field $A_{2}$ is unconstrained so can fluctuate at the boundary.
We will call the set of boundary conditions (\ref{asusybc3a1}) - (\ref{asusybc3b0}) \textit{magnetic-like} A-type boundary conditions.

\end{itemize}

\subsubsection*{Boundary degree of freedom for the A-type from the bulk vector multiplet}
We see that two sets of the A-type boundary conditions,  
(\ref{asusybc3c})-(\ref{asusybc3e}) and 
(\ref{asusybc3a1})-(\ref{asusybc3b}),
provide decomposition of the 3d $\mathcal{N}=4$ vector multiplet $V_{\mathcal{N}=4}$ under $\mathcal{N}=(2,2)$ supersymmetry at the boundary.
The two-dimensional gauge field $A_m$
and the two real scalar fields $\phi^a$, $a=3,4$, which form a complex scalar field, 
are naturally packaged into a 2d $\mathcal{N}=(2,2)$ vector multiplet $V^{(2,2)}$ 
or field strength multiplet $\Sigma^{(2,2)}$.
Meanwhile, from two real scalar fields $A_2$ and $\phi^5$, 
one can form a 2d $\mathcal{N} = (2,2)$ twisted chiral multiplet $\widetilde{\Sigma}^{(2,2)}$, 
which is charged under axial $U(1)_C$ R-symmetry group. 
We let $\rho$ be the dual photons 
defined by $\frac12 \epsilon_{\mu\nu\rho}F^{\nu\rho}=\partial_{\mu}\rho$ 
for each of the Abelian factors of gauge group where $A_{2}$, 
which is surviving degree of freedom when considering Dirichlet-like boundary condition, 
appears in the LHS.
Then $\widetilde{\rho} = \phi^5 + i \rho$ is charged under $U(1)_C$ 
and becomes a scalar component of twisted chiral multiplet.
Therefore, 3d $\mathcal{N}=4$ vector multiplet $V_{\mathcal{N}=4}$ 
can be decomposed into 2d $\mathcal{N}=(2,2)$ vector multiplet $V^{(2,2)}$, 
or field strength multiplet, which is a twisted chiral multiplet $\Sigma^{(2,2)}$, 
and 2d $\mathcal{N}=(2,2)$ twisted chiral multiplet 
$\widetilde{\Sigma}^{(2,2)}$;
\begin{align}
\label{adec1}
V_{\mathcal{N}=4}
&\rightarrow (V^{(2,2)}, \widetilde{\Sigma}^{(2,2)}).
\end{align}

The 3d $\mathcal{N}=4$ supersymmetric parameters consist of two copies of the 3d $\mathcal{N}=2$ supersymmetric parameters, $\xi_{1}$ and $\xi_{2}$. 
The projection (\ref{asusybc1}) admits two right-moving supersymmetric parameters and two left-moving supersymmetric parameters.\footnote{
We denote the right-moving fermion by $\Psi^{+}$ and the left-moving fermion by $\Psi^{-}$,
\begin{align}
\label{2dfer1}
\gamma^{2}\Psi^{+}&=\Psi^{+},& 
\gamma^{2}\Psi^{-}&=-\Psi^{-}. 
\end{align}
One raises and lowers the spinor indices by the antisymmetric tensor 
$\Psi_{\alpha}=
\epsilon_{\alpha\beta}
\Psi^{\beta}$ with $\epsilon_{+-}=1$ so that 
$\Psi^{-}=\Psi_{+}$, $\Psi^{+}=-\Psi_{-}$. }
Denoting the complex supersymmetric parameters of 2d $\mathcal{N}=(2,2)$ supersymmetry as
\begin{align}
\label{22susy1a}
\xi^{+}&:=\frac12 (\xi_{1}^+ + \overline{\xi}_{1}^+) + \frac{i}{2}(\xi_{2}^+ + \overline{\xi}_{2}^+)	\, 	,	& 
\xi^{-}&:=\frac{1}{2i} (\xi_{1}^- - \overline{\xi}_{1}^-) + \frac12(\xi_{2}^- - \overline{\xi}_{2}^-)	\,	, 
\end{align}
with
$\overline{\xi}^{+}=(\xi^{+})^{*}$ and $\overline{\xi}^{-}=(\xi^{-})^{*}$ at the boundary,
the axial $U(1)_A$ and the vector $U(1)_V$ of them may take
\begin{align}
\label{22susy1b}
\begin{array}{c|c|c|c}
&SO(1,1)&U(1)_{A}&U(1)_V\\ \hline
\xi^{+}&+&-&+\\
\xi^{-}&-&+&+\\
\overline{\xi}^{+}&+&+&-\\
\overline{\xi}^{-}&-&-&-\\
\end{array}
\end{align}

For the the electric-like A-type boundary conditions, which allow both left-moving and right-moving fermions,
we similarly denote the two-dimensional fermionic fields by
\begin{align}
\label{22susy2a}
\lambda^{+}&:=\frac12\left(\lambda+\overline{\lambda}\right)
+\frac{i}{2}\left(\psi_{\phi}+\overline{\psi}_{\phi}\right),& 
\lambda^{-}&:=\frac{1}{2i}\left(\lambda-\overline{\lambda}\right)+\frac12\left(\psi_{\phi}-\overline{\psi}_{\phi}\right),
\end{align}
with 
$\overline{\lambda}^{+}=(\lambda^+)^{\dag}$ and $\overline{\lambda}^{-}=(\lambda^{-})^{\dag}$ at the boundary. 
Their R-charges would be
\begin{align}
\label{22susy2b}
\begin{array}{c|c|c|c}
&SO(1,1)&U(1)_{A}&U(1)_V\\ \hline
\lambda^{+}&+&+&+\\
\lambda^{-}&-&-&+\\
\overline{\lambda}^{+}&+&-&-\\
\overline{\lambda}^{-}&-&+&-\\
\end{array}
\end{align}
We write a complex scalar field as $\phi:=\phi^{1}+i\phi^{2}$. 
Given the notation above, supersymmetric transformation laws of the component fields 
$(A_{m}, \phi, \lambda^{+}, \lambda^{-}, \overline{\lambda}^{+}, \overline{\lambda}^{-})$, 
which form an $\mathcal{N}=(2,2)$ vector multiplet, would be \cite{Witten:1993yc, Hori:2003ic}
\begin{align}
\label{22susyvm1}
\delta A_{\pm}&=i\overline{\xi}_{\pm}\lambda_{\pm}+i\xi_{\pm}\overline{\lambda}_{\pm},\\
\label{22susyvm2}
\delta \phi&=-i\overline{\xi}_{+}\lambda_{-}-i\xi_{-}\overline{\lambda}_{+},\\
\label{22susyvm3}
\delta \lambda_{+}&=2\partial_{+}\overline{\phi}\xi_{-}+(iD-v_{01})\xi_{+},\\
\label{22susyvm4}
\delta \lambda_{-}&=2\partial_{-}\overline{\phi}\xi_{+}+(iD+v_{01})\xi_{-},\\
\label{22susyvm5}
\delta D&=-\overline{\xi}_{+}D_{-}\lambda_{+}-\overline{\xi}_{-}D_{+}\lambda_{-}
+\xi_{+}D_{-}\overline{\lambda}_{+}+\xi_{-}D_{+}\overline{\lambda}_{-},
\end{align}
where $D$ is an auxiliary field, which is expressed as some function of $\phi$ 
where the detail form of it can be determined once the detail of coupling to the boundary fields is given.

The magnetic-like A-type boundary conditions 
$\gamma^{2}\lambda=-\overline{\lambda}$, $\gamma^{2}\psi_{\phi}=-\overline{\psi}_{\phi}$ also yield  
both left-moving and right-moving fermions. 
We similarly denote the two-dimensional fermions by 
\begin{align}
\label{22susy1c}
\chi^{+}&:=\frac{1}{2i}\left(\lambda-\overline{\lambda}\right)+\frac12\left(\psi_{\phi}-\overline{\phi}_{\phi}\right),& 
\chi^{-}&:=\frac12\left(\lambda+\overline{\lambda}\right)+\frac{i}{2}\left(\psi_{\phi}+\overline{\phi}_{\phi}\right),
\end{align}
and their complex conjugate, 
$\overline{\chi}^{+}=(\chi^{+})^{\dag}$, $\overline{\chi}^{-}=(\chi^{-})^{\dag}$. 

They would carry the R-charges as
\begin{align}
\label{22susy1d}
\begin{array}{c|c|c|c}
&SO(1,1)&U(1)_{A}&U(1)_V\\ \hline
\chi^{+}&+&+&+\\
\chi^{-}&-&-&+\\
\overline{\chi}^{+}&+&-&-\\
\overline{\chi}^{-}&-&+&-\\
\end{array}
\end{align}
The twisted chiral multiplet $\widetilde{\Sigma}^{(2,2)}$ has the component fields 
$(\widetilde{\rho}, \chi^{+}, \chi^{-}, \overline{\chi}^{+}, \overline{\chi}^{-})$. 
The supersymmetry transformation laws would take the form \cite{Witten:1993yc, Hori:2003ic}
\begin{align}
\label{22susytcm1}
\delta \widetilde{\rho}&=\overline{\xi}_{+}\chi_{-}-\xi_{-}\overline{\chi}_{+},\\
\label{22susytcm2}
\delta \overline{\chi}_{+}&=2i\partial_{+}\widetilde{\rho}\overline{\xi}_{-}+G\overline{\xi}_{+},\\
\label{22susytcm3}
\delta \chi_{-}&=-2i\partial_{-}\widetilde{\rho}\overline{\xi}_{+}+G\overline{\xi}_{-},\\
\label{22susytcm4}
\delta G&=-2i\xi_{+}\partial_{-}\overline{\chi}_{+}-2i\overline{\xi}_{-}\partial_{+}\chi_{-},
\end{align}
where $G$ is some function of $\widetilde{\rho}$.

\subsubsection{B-type boundary conditions}
\label{secvmb}

Next we consider the B-type conditions where the projection condition on the supersymmetric parameter $\xi$ is
\begin{align}
\label{bsusybc1}
\gamma^{2}\xi&= -\xi  \, .
\end{align}
Here and in the followings, we choose a convention that 
this gives the right-moving supercharges, 
which leads to the chiral $\mathcal{N}=(0,4)$ supersymmetry at the two-dimensional boundary.\footnote{In this convention, the projection condition for the supersymmetric parameter 
$\gamma^2 \xi = \xi$ preserves $\mathcal{N}=(4,0)$ supersymmetry at the boundary.}

Applying the ansatz 
\begin{align}
\label{bsusybc2}
\gamma^{2} \lambda = e^{2i\theta} \lambda	\,	,	\qquad
\gamma^{2} \psi_\phi = e^{2i\theta} \psi_\phi
\end{align}
for the fermionic boundary conditions, which doesn't change the equations of motion when $\theta=0$ or $\frac{\pi}{2}$, we can find the bosonic boundary conditions. 
These two boundary conditions for the fermionic fields determine their chiralities at the boundary. 
When $\theta=0$ (resp. $\theta=\frac{\pi}{2}$), the associated two-dimensional fermions are right-moving (resp. left-moving).

\begin{itemize}[leftmargin=5mm]
\item[i)] $\gamma^2 \lambda = - \lambda$ and $\gamma^2 \psi_{\phi} = - \psi_{\phi}$ ($\theta = \frac{\pi}{2}$)

With this choice of the fermionic boundary condition, 
it follows that
\begin{align}
\begin{split}
\overline{\xi} \lambda	=	0	\,	,	\quad
\overline{\xi} \sigma^2 \lambda	=	0	\,	,	\quad
\overline{\xi} \overline{\psi}_{\phi}	=	0	\,	,	\quad	
\overline{\xi} \sigma^2 \overline{\psi}_{\phi}	=	0
\end{split} \label{b1r}
\end{align}
so the general boundary conditions (\ref{susybc1a}) turn into 
\begin{align}
\label{bsusybc3aa}
\begin{split}
0&=\frac{1}{2} (i F_{{m} 2} - D_{m} \sigma) (\overline{\xi} \sigma^{{m}} \lambda) 
+ \frac{1}{2} (i F^{{m} 2} + D_{m} \sigma) (\xi \sigma^{{m}} \overline{\lambda}) 		\\
&-\frac{1}{\sqrt{2}} D_{m} \phi (\overline{\xi} \sigma^{{m} 2} \overline{\psi}_{\phi})
+\frac{1}{\sqrt{2}} D_{m} \overline{\phi} (\xi \sigma^{{m} 2} \psi_{\phi})		\,	. 
\end{split}
\end{align}
Therefore, with identification \eqref{iden}, we find
\begin{align}
\label{bsusybc3a}
F_{2m}&=0	\,	,	\\
\label{bsusybc3b}
D_{m}\phi^{i}&=0	\,	.
\end{align}
The first condition (\ref{bsusybc3a}) is the Neumann-like boundary condition for the two-dimensional gauge field $A_{m}$,  
while the second condition (\ref{bsusybc3b}) is the Dirichlet-like boundary condition for the three scalar fields $\phi^{i}$. 
The condition (\ref{bsusybc3a}) can be rephrased as the Dirichlet-like boundary conditions for the scalar field $A_{2}$. 
We call this set of boundary conditions (\ref{bsusybc3a}) and (\ref{bsusybc3b}) the \textit{electric-like} B-type boundary conditions.

\item[ii)] $\gamma^2 \lambda = \lambda$ and $\gamma^2 \psi_{\phi} = \psi_{\phi}$ ($\theta = 0$)

Choosing $\theta=0$ for the fermionic boundary conditions in (\ref{bsusybc2}), we get 
\begin{align}
\label{b2r}
\overline{\xi} \sigma_{m} \lambda		=	0	\,	,	\quad
\overline{\xi} \sigma^{{m} 2} \lambda		=	0	\,	,	\quad
\overline{\xi} \sigma^{{m} 2} \overline{\psi}_{\phi}		=	0	\,	.
\end{align}
The generic boundary conditions (\ref{susybc1a}) then reduce to 
\begin{align}
\begin{split}
0&=
\frac{1}{2} (-i F_{01} + D^2 \sigma + [\phi, \overline{\phi}]) (\overline{\xi} \lambda)		
+\frac{1}{2} (-i F_{01} - D^2 \sigma - [\phi, \overline{\phi}]) (\xi \overline{\lambda})		\\
&
\quad +\frac{1}{\sqrt{2}} (D^2 \phi + [\sigma, \phi] ) (\overline{\xi} \overline{\psi}_{\phi})		
- \frac{1}{\sqrt{2}} (D^2 \overline{\phi} - [\sigma, \overline{\phi}]) (\xi \psi_{\phi}) 		
\end{split}	\label{bsusybc3cc}
\end{align}
Thus, 
with identification \eqref{iden},
one finds
\begin{align}
\label{bsusybc3c}
&F_{01}=0 	\,	,\\
\label{bsusybc3d}
&D_2 \phi^i -\frac{1}{2} i \epsilon^{ijk} [\phi^j, \phi^k]	=	0
\end{align}
where $\epsilon^{ijk}$ is the Levi-Civita symbol with $\epsilon^{345}=1$. 
The first condition (\ref{bsusybc3c}) is the Dirichlet-like condition for the two-dimensional gauge field $A_{m}$. 
The scalar field $A_{2}$ is unconstrained and can fluctuate at the boundary.
Note that the boundary condition for three scalar fields $\phi^{i}$ is not Neumann-like but rather they satisfy Nahm-like equations.
They originate from the existence of fluctuating $A_2$ at the boundary \cite{Gaiotto:2008sa}. 
We will call this set of boundary conditions (\ref{bsusybc3c}) and (\ref{bsusybc3d}) the \textit{magnetic-like} B-type boundary conditions.

\end{itemize}

\subsubsection*{Boundary degree of freedom for the B-type from the bulk vector multiplet}

We can also see 
the two sets of the B-type conditions (\ref{bsusybc3a})-(\ref{bsusybc3b}) and (\ref{bsusybc3c})-(\ref{bsusybc3d})
provide the decomposition of the 3d $\mathcal{N}=4$ vector multiplet 
under the preserved $\mathcal{N}=(0,4)$ supersymmetry at the boundary.
We observed that 
for the electric-like B-type boundary conditions
the two-dimensional gauge field $A_{m}$ can fluctuate and
a pair of left-moving fermions transforming as $(\bm{2,}\bm{2})_{-}$ survive at the boundary.
They are part of the 2d $\mathcal{N}=(0,4)$ vector multiplet $V^{(0,4)}$, which also contains 
an auxiliary field transforming as $(\bm{1},\bm{3})_{0}$ that 
originate from the auxiliary field $F$ in the 3d $\mathcal{N}=4$ vector multiplet (see (\ref{3dn4vm1a})). 
On the other hand, for the magnetic-like B-type boundary conditions, 
the scalar fields $\phi^{i}$ and $A_{2}$ can fluctuate at the boundary and can be combined into 
the two complex scalar fields transforming as $(\bm{2},\bm{1})_{0}$.
Also a pair of right-moving fermions $(\bm{1},\bm{2})_{+}$ survive at the boundary.
Therefore, they form the $\mathcal{N} = (0,4)$ twisted hypermultiplets $\widetilde{H}^{(0,4)}$. 
Hence for the B-type conditions, the 3d $\mathcal{N}=4$ vector multiplet $V_{\mathcal{N}=4}$ splits into two parts; 
\begin{align}
\label{bdec1}
V_{\mathcal{N}=4}&\rightarrow 
(V^{(0,4)}, \widetilde{H}^{(0,4)}).
\end{align}

The projection (\ref{bsusybc1}) reduces two copies $\xi_1$, $\xi_2$ of 3d $\mathcal{N}=2$ supersymmetric parameters to four real left-moving supersymmetric parameters. 
We write them as $\xi^{A\dot{A}}$ where the indices $A,B,\cdots=1,2$ transform as a doublet under $SU(2)_C$ while the indices $\dot{A} ,\dot{B}, \cdots = \dot{1}, \dot{2}$ transform as a doublet under $SU(2)_H$. 
We denote the four supersymmetric parameters of 2d $\mathcal{N}=(0,4)$ supersymmetry by
\begin{align}
\label{04susy}
\xi^{-1\dot{1}}	&	:=	\xi_{1}^-	\,	,	& 
\xi^{-1\dot{2}}	&	:=	-\xi_{2}^-	\,	,	& 
\xi^{-2\dot{1}}	&	:=	\overline{\xi}_{2}^-	\,	,	& 
\xi^{-2\dot{2}}	&	:=	\overline{\xi}_{1}	\,	.
\end{align}
The electric-like B-type boundary conditions $\gamma^{2}\lambda=-\lambda$, $\gamma^{2}\psi_{\phi}=-\psi_{\phi}$ lead to  left-moving fermions. 
We take them as the doublets under the both $SU(2)_C$ and $SU(2)_H$ so that 
\begin{align}
\label{04susyvm1}
\lambda^{-1\dot{1}}	&	:=	\lambda^-	\,	,	& 
\lambda^{-1\dot{2}}	&	:=	-\psi_\phi^-	\,	,	& 
\lambda^{-2\dot{1}}	&	:=	\overline{\psi}_{\phi}^-	\,	,	& 
\lambda^{-2\dot{2}}	&	:=	\overline{\lambda}^-	\,	.
\end{align}
Denoting the component fields for the vector multiplet $V^{(0,4)}$ by $(A_{m}, \lambda^{-A\dot{A}} )$, the supersymmetry transformation\footnote{From 3d $\mathcal{N}=4$ supersymmetry transformation with projection and boundary conditions, we can see
\begin{align}
\delta (A_0 - A_1) 		&=	2 i \xi_{A\dot{A}}^{-} \lambda^{-A\dot{A}}	\\
\delta (A_0 + A_1) 		&=	0		\\
\delta \lambda^{-A\dot{B}}	&=	i D^{A}_{\ C} \xi^{-C\dot{B}}  + F_{01} \xi^{-A \dot{B}}		\,	,		
\end{align}
at the boundary.
Here, $D^{A}_{\ B} = \frac{1}{2} [ \widetilde{X}^{AY'}, \widetilde{X}_{BY'} ]$ with $\widetilde{X}^{AY'} = \begin{pmatrix} \sigma - i A_2 & -\sqrt{2}\phi \\ \sqrt{2}\overline{\phi} & \sigma + i A_2 \end{pmatrix}$, which is a scalar component of twisted hypermultiplet, 
where indices $Y' =1',2'$ denote the doublet under $SU(2)^{'}_F$ global symmetry. 
}
would be 
\begin{align}
\label{04susyvm2}
\delta A_{-}	&=	2 i \xi^{-}_{A\dot{A}} \lambda^{-A\dot{A}}	\,	,	\\
\label{04susyvm3}
\delta \lambda^{-A\dot{B}}	&=	i D^{A}_{\ C} \xi^{-C\dot{B}}  + F_{01} \xi^{-A \dot{B}}		\,	,		
\end{align}
where $D^{A}_{\ B}$ would be some function of scalar field $\widetilde{X}^{AY'}$ in the twisted hypermultiplet for generic coupling to boundary fields. We use the antisymmetric tensor $\epsilon_{AB}$ and $\epsilon_{\dot{A} \dot{B}}$ with $\epsilon_{+-} = \epsilon_{\dot{+}\dot{-}} = 1$ to raise or lower indices $A, B, \ldots$ and $\dot{A}, \dot{B}, \ldots$, respectively.

The magnetic-like B-type boundary conditions $\gamma^{2}\lambda=\lambda$, $\gamma^{2}\psi_{\phi}=\psi_{\phi}$ lead to right-moving fermions. 
We can take them as a doublet under the $SU(2)_{H}$ and also a doublet under the additional global symmetry $SU(2)_{F}'$.
We write them as $\widetilde{\Psi}^{+\dot{A}Y'}$ 
\begin{align}
\widetilde{\Psi}^{+\dot{A}Y'} = \begin{pmatrix} \overline{\chi}^+ & \lambda^+ \\ \overline{\lambda}^+ & - \chi^+ \end{pmatrix}
\end{align}
where the indices $Y' = 1' , 2'$ represent the doublet under the $SU(2)_F'$. 
The supersymmetry transformations of the component fields $(\widetilde{X}^{AY'},\widetilde{\Psi}^{+A'Y'})$, which form a twisted hypermultiplet, can also be obtained from 3d $\mathcal{N}=4$ supersymmetry transformation with projection and boundary conditions, 
\begin{align}
\label{04susythm1}
\delta \widetilde{X}^{AY'}			&=	-2 \xi^{-A\dot{B}} \epsilon_{\dot{B}\dot{C}} \widetilde{\Psi}^{+\dot{C} Y'}	\,	,	\\
\label{04susythm2}
\delta \widetilde{\Psi}^{+\dot{A} Y'}	&=	-i \xi^{-B\dot{A}} \epsilon_{BC} (\partial_0 + \partial_1) \widetilde{X}^{CY'}	\,	, 
\end{align}
which is the supersymmetry transformation of the $\mathcal{N}=(0,4)$ twisted hypermultiplet 
where Dirichlet-like condition $A_0=A_1=0$ is incorporated. 
See also \cite{Witten:1994tz, Gates:1994bu}.

\subsubsection{Reduction from the extended Bogomolny equation}
\label{secextended}

The BPS equations of the topologically twisted 4d $\mathcal{N}=4$ SYM theories on a 4-manifold $M_{4}$ 
have been studied in \cite{Kapustin:2006pk}, which read
\begin{align}
\label{gltwist1a}
(F-\phi\wedge \phi+t d_{A}\phi)^{+}&=0,\\
\label{gltwist1b}
(F-\phi\wedge \phi-t^{-1}d_{A}\phi)^{-}&=0,\\
\label{gltwist1c}
d_{A}\star \phi&=0,
\end{align}
where $A$ is a four-dimensional anti-hermitian gauge field that is a connection 
on a $G$-bundle $E\rightarrow M_{4}$, 
and $\phi$ is a bosonic one-form field valued in anti-hermitian matrix given the adjoint representation of the Lie algebra of $G$. 
$d_{A}=d+[A,\cdot]$ is the covariant exterior derivative, 
$F=dA+A\wedge A$ is the field strength, 
$\star$ is the Hodge star operator, 
and $t$ is a real constant parametrizing a family of topological twisted theories. 
Especially when $t=1$, the set of equations (\ref{gltwist1a})-(\ref{gltwist1c}) can be written as 
\begin{align}
\label{gltwist1d}
F-\phi\wedge \phi+\star d_{A}\phi&=0,\\
\label{gltwist1e}
d_{A}\star \phi&=0.
\end{align}

Equation (\ref{gltwist1d}) is called the extended Bogomolny equation in \cite{Kapustin:2006pk}. 
It has been argued that the BPS equations (\ref{gltwist1d}) together with (\ref{gltwist1e}) provide a various family of the BPS equations in lower dimensions by performing the reduction on a given $M_4$, \textit{e.g.} on $M_{4}=C\times \Sigma$ where $C$ and $\Sigma$ are Riemann surfaces \cite{Bershadsky:1995vm,Kapustin:2006pk}, and on $M_{4}=M_{3}\times \mathbb{R}_{+}$ where $M_{3}$ is a 3-manifold and $\mathbb{R}_{+}$ is a half line \cite{Gaiotto:2011nm,Witten:2011zz}. 
Here, we would like to see our BPS boundary conditions for the 3d $\mathcal{N}=4$ vector multiplet in the reduction of the extended Bogomolny equation.

We consider the equations on a 4-manifold $M_{4}=\mathbb{R}_{+}\times M_{3}$. 
We express the gauge field as $A=A_{0}dx^{0}+\widetilde{A}$, and the one-form as $\phi=\phi_{0}dx^{0}+\widetilde{\phi}$ 
where $x^{0}$ is the coordinate on the half line $\mathbb{R}_{+}$. 
Taking the $x^{0}$ independent parts from (\ref{gltwist1d})-(\ref{gltwist1e}), 
one obtains the BPS equations on $M_{3}$
\begin{align}
\label{gltwist2a}
\widetilde{F}-\widetilde{\phi}\wedge \widetilde{\phi}
&=\star\left(d_{\widetilde{A}}\phi_{0}-[A_{0},\widetilde{\phi}] \right),\\
\label{gltwist2b}
d_{\widetilde{A}}A_{0}+[\phi_{0},\widetilde{\phi}]&=\star d_{\widetilde{A}}\widetilde{\phi},\\
\label{gltwist2c}
d_{\widetilde{A}}^{*}\widetilde{\phi}+[A_{0},\phi_{0}]&=0,
\end{align}
where the exterior derivative $d_{\widetilde{A}}$,  
the Hodge operator $\star$ and 
$d_{\widetilde{A}}^*=\star d_{\widetilde{A}}\star$ are defined on the 3-manifold 
$M_{3}$. 
We further take $M_{3}=\mathbb{R}_{+}\times C$ 
and write $\widetilde{A}=A_{2}dx^{2}+A_{z}dz+A_{\overline{z}}d\overline{z}$, 
$\widetilde{\phi}=\phi_{2}dx^{2}+\phi_{z}dz+\phi_{\overline{z}}d\overline{z}$ 
where $\mathbb{R}_{+}$ is the half line $x^{2}\ge 0$ 
and $z,\overline{z}$ are the local complex coordinates on the Riemann surface $C$. 
By squaring (\ref{gltwist2a})-(\ref{gltwist2c}) and integrating by parts, 
one finds that $A_{0}=\phi_{2}=0$. 
Let us denote the metric on 3-manifold $M_{3}$ 
by $ds^{2}=(dx^{2})^{2}+2|dz|^{2}$ 
and choose a gauge in which $A_{2}=0$. 
Then (\ref{gltwist2a})-(\ref{gltwist2c}) 
are now simplified 
to \cite{Kapustin:2006pk}
\begin{align}
\label{gltwist3a}
F_{z\overline{z}}-[\phi_{z},\phi_{\overline{z}}]&=i\partial_{2}\phi_{0},\\
\label{gltwist3b}
\partial_{2}A_{\overline{z}}&=-iD_{\overline{z}}\phi_{0},\\
\label{gltwist3c}
i[\phi_{0},\phi_{z}]&=\partial_{2}\phi_{z},\\
\label{gltwist3d}
D_{z}\phi_{\overline{z}}&=0.
\end{align}
As a 4-manifold is now a product space $M_{4}=\mathbb{R}_{+}\times \mathbb{R}_{+}\times C$, the topological twisting is not performed on the 4-manifold but on the two-dimensional surface $C$. 
When $C=\mathbb{R}^2$, which we will consider, the above configuration on a 3-manifold $M_{3}=\mathbb{R}_{+}\times C$ with a boundary at $x^{2}=0$ may admit maximally four supercharges.
Regarding boundary conditions, given a field, it is reasonable to expect that there is either normal derivative or tangential derivative of it but not both in the (BPS) boundary conditions or equations that the boundary fields should satisfy.
So by picking sets of equations among \eqref{gltwist3a}-\eqref{gltwist3d}, more precisely, one in \eqref{gltwist3a} or \eqref{gltwist3b} and one in \eqref{gltwist3c} or \eqref{gltwist3d}, and by taking terms in equations to be separately zero, we may be able to find four consistent sets of BPS boundary conditions we are interested in.
Meanwhile, we note that \eqref{gltwist3a} and \eqref{gltwist3b} have terms relevant to the Dirichlet-like and Neumann-like boundary conditions for the two-dimensional gauge fields $A_{z}$, $A_{\overline{z}}$, respectively, whereas (\ref{gltwist3c}) and (\ref{gltwist3d}) contain the Neumann-like and Dirichlet-like boundary conditions for the one-form fields $\phi_{z}$, $\phi_{\overline{z}}$, respectively.

In order to see our boundary conditions in the reduced extended Bogomolny equations \eqref{gltwist3a}-\eqref{gltwist3d},
we take $C=\mathbb{R}^{2}$ where the one-form fields $\phi_{z}$ and $\phi_{\overline{z}}$ reduce to the scalar fields 
and we set
\begin{align}
\label{gltwist4a}
\partial_{z}&=\frac{1}{\sqrt{2}}
(\partial_{\hat{0}}-i\partial_{\hat{1}}),& 
\partial_{\overline{z}}&=\frac{1}{\sqrt{2}}
(\partial_{\hat{0}}+i\partial_{\hat{1}}),\\
\label{gltwist4b}
A_{z}&=\frac{1}{\sqrt{2}}(A_{\hat{0}}-iA_{\hat{1}}),&
A_{\overline{z}}&=\frac{1}{\sqrt{2}}(A_{\hat{0}}+iA_{\hat{1}}),\\
\label{gltwist4c}
\phi_{0}&=\phi^{5}, & 
\phi_{z}&=\frac{1}{\sqrt{2}}(\phi^{3}-i\phi^{4}),& 
\phi_{\overline{z}}&=\frac{1}{\sqrt{2}}
(\phi^{3}+i\phi^{4}),
\end{align} 
where $m, n = \hat{0}, \hat{1}$ are space-time indices on $\mathbb{R}^{2}$ while $i,j,\cdots=3,4,5$ and $a,b,\cdots=3,4$ label the scalar fields.

\begin{itemize}[leftmargin=5mm]

\item[A-i)] 
From \eqref{gltwist3b} and (\ref{gltwist3c}) 

By taking both of the LHS and the RHS of all the equations to be separately zero, we have
\begin{align}
F_{2 m} = 0 	\, 	,	\qquad
D_{m}\phi^{5} = 0 	\,	 , 	\qquad
D_{2}\phi^{a} = 0 	\, 	,	\qquad
[\phi^{5},\phi^{a}] = 0 \,	,
\end{align}
and one can identify them with the electric-like A-type conditions (\ref{asusybc3c})-(\ref{asusybc3e}).

\item[A-ii)] From (\ref{gltwist3a}) and (\ref{gltwist3d}) 

We can obtain
\begin{align}
F_{\hat{0}\hat{1}} = 0 	\, ,	\qquad
D_{2}\phi^{5} = 0 	\, ,		\qquad
[\phi^{a},\phi^{b}] = 0 	\, ,	\qquad
D_{m}\phi^{a} = 0 	\, 	
\end{align}
by setting every term in (\ref{gltwist3a}) to be zero.
These are the magnetic-like A-type conditions (\ref{asusybc3a1})-(\ref{asusybc3b0}).

\item[B-i)] From (\ref{gltwist3b}) and (\ref{gltwist3d})

Similarly, by taking both the LHS and the RHS in (\ref{gltwist3b}) to be separately zero, we get
\begin{align}
F_{2 m} = 0 	\, ,	\qquad
D_{m}\phi^{i} = 0 	\, 	.
\end{align}
This set of equations are the electric-like B-type conditions (\ref{bsusybc3a})-(\ref{bsusybc3b}).

\item[B-ii)] From (\ref{gltwist3a}) and (\ref{gltwist3c}) 

We can obtain
\begin{align}
F_{mn} = 0 	\, ,	\qquad
D_{2}\phi^{i}+\epsilon^{ijk}[\phi^{j},\phi^{k}] = 0 	\, 	
\label{b-ii-eq}
\end{align}
by taking terms in \eqref{gltwist3a} to be zero after arrangement,
where we have restored the gauge fixed value $A_{2}=0$.  
Taking into account that $\phi^i$'s are anti-hermitian here, we see that both equations are the magnetic-like B-type conditions (\ref{bsusybc3c})-(\ref{bsusybc3d}).\footnote{\eqref{bsusybc3d} can be recovered from \eqref{b-ii-eq} via $A_2 \rightarrow -i A_2$ and $\phi^j \rightarrow -\frac{1}{2} i \phi^j$.}

\end{itemize}

\subsection{Hypermultiplets}
\label{sechm}

The 3d $\mathcal{N}=4$ hypermultiplets contain 
complex scalar fields $\bm{q}$ and fermionic fields $\bm{\psi}$ transforming as
\begin{align}
\begin{split}
\bm{q}			&: (\bm{1},\bm{1},\bm{2})			\\
\bm{\psi}			&: (\bm{2},\bm{2},\bm{1})
\end{split}\label{3dn4hm1a}
\end{align}
under the $SO(1,2)\times SU(2)_{C}\times SU(2)_{H}$.

Also, the 3d $\mathcal{N}=4$ hypermultiplets in representation $R$ of the gauge group can be expressed as a combination of the two 3d $\mathcal{N}=2$ chiral multiplets 
$Q(q,\psi,F_{q})$ and $\widetilde{Q}(\widetilde{q},\widetilde{\psi},F_{\widetilde{q}})$ 
transforming in conjugate representations, $R$ and $\overline{R}$, of the gauge group. 
The action of the 3d $\mathcal{N}=4$ hypermultiplets coupled to 3d $\mathcal{N}=4$ vector multiplet is given by 
\begin{align}
\label{3dn4hm1b}
S&=S_{K}^{\mathcal{N}=2}+S_{W}^{\mathcal{N}=2}
\end{align}
where 
\begin{align}
\label{3dn4hm1c}
S_{K}^{\mathcal{N}=2}
&=-\int d^{3}x d^{4}\theta\  
\left(
\overline{Q}e^{-2V}Q
+\overline{\widetilde{Q}}e^{-2V} \widetilde{Q}
\right)
\end{align}
is the kinetic terms and
\begin{align}
\label{3dn4hm1d}
S_{W}^{\mathcal{N}=2}
= -\sqrt{2}i\int d^{3}x d^{2}\theta  \left( \widetilde{Q}\Phi Q \right)
+c.c \, . 
\end{align}
is the superpotential terms and $c.c.$ stands for the complex conjugate.

In terms of the component fields, the action (\ref{3dn4hm1c}) can be expressed as 
\begin{align}
\begin{split}
\label{3dn4hm1c1}
S_{K}^{\mathcal{N}=2}
=\int d^{3}x 
\Biggl[
&-D_{\mu}\overline{q}D^{\mu}q
-i\overline{\psi}\sigma^{\mu}D_{\mu}\psi
+\overline{F}_{q}F_{q}
-i \overline{\psi} \sigma \psi
-\sqrt{2}i  \overline{\psi} \overline{\lambda} q
-\sqrt{2}i  \overline{q} \lambda \psi
-\overline{q} D q - \overline{q} \sigma^{2} q  \\
&
-D^{\mu}\widetilde{q} D_{\mu}\overline{\widetilde{q}}
-i\widetilde{\psi}\sigma^{\mu}D_{\mu}\overline{\widetilde{\psi}}
+F_{\widetilde{q}} \overline{F}_{\widetilde{q}}
+i \widetilde{\psi} \sigma \overline{\widetilde{\psi}}  
+\sqrt{2} i \widetilde{q} \overline{\lambda} \overline{\widetilde{\psi}} 
+\sqrt{2} i \widetilde{\psi} \lambda \overline{\widetilde{q}}  
+\overline{\widetilde{q}} D \widetilde{q}
-\widetilde{q} \sigma^{2} \overline{\widetilde{q}} 
\Biggr]
\end{split}
\end{align}
where $\sigma = \sigma^a T^a_R, \, D = D^a T^a_R, \, \lambda = \lambda^a T^a_R, \, \overline{\lambda} = \overline{\lambda}^a T^a_R$ 
and the action (\ref{3dn4hm1d}) as
\begin{align}
\label{3dn4hm1c2}
S_{W}^{\mathcal{N}=2} = 
\int d^{3}x\ \Biggl[ 
-\sqrt{2}i \left( F_{\widetilde{q}} \phi q + \widetilde{q} F_{\phi} q + \widetilde{q} \phi F_{q} \right)
+ i \sqrt{2} \left( \widetilde{\psi} \psi_{\phi} q + \widetilde{q} \psi_{\phi} \psi + \widetilde{\psi} \phi \psi \right)
\Biggr]
+c.c. 
\end{align}
where the covariant derivatives are defined by 
\begin{align}
\begin{split}
D_{\mu}q&=\partial_{\mu}q-iA_{\mu}q,							\qquad		D_{\mu}\overline{q}=\partial_{\mu}\overline{q}+i\overline{q}A_{\mu},	\\
D_{\mu}\widetilde{q}&=\partial_{\mu}\widetilde{q}+i\widetilde{q}A_{\mu},	\qquad		D_{\mu}\overline{\widetilde{q}}=\partial_{\mu}\overline{\widetilde{q}}-iA_{\mu}\overline{\widetilde{q}} \, .
\end{split}\label{covderiv0}
\end{align}

The actions (\ref{3dn4hm1b}) is invariant under the supersymmetry transformations
\begin{align}
\label{3dn2s1a}
\delta q			&= \sqrt{2}\xi \psi	\,	,	\\
\label{3dn2s1b}
\delta \widetilde{q}	&=\sqrt{2}\xi\widetilde{\psi}	\,	,	\\
\label{3dn2s1c}
\delta\psi			&=
\sqrt{2}i\gamma^{\mu}\overline{\xi} D_{\mu}q 
+\sqrt{2}\xi F_{q}-\sqrt{2}i\overline{\xi}\sigma q	\,	,	\\
\label{3dn2s1d}
\delta\widetilde{\psi}	&=
\sqrt{2}i \gamma^{\mu}\overline{\xi} D_{\mu}\widetilde{q}
+\sqrt{2}\xi F_{\widetilde{q}}+\sqrt{2}i \overline{\xi}\sigma \widetilde{q}	\,	,	\\
\label{3dn2s1e}
\delta F_{q}		&=
\sqrt{2}i\overline{\xi}\sigma^{\mu}D_{\mu}\psi
+2i(\overline{\xi}\overline{\lambda})q+\sqrt{2}i(\overline{\xi}\psi)\sigma	\,	,	\\
\label{3dn2s1f}
\delta F_{\widetilde{q}}	&=
\sqrt{2}i \overline{\xi}\sigma^{\mu}D_{\mu}\widetilde{\psi}
-2i(\overline{\xi}\overline{\lambda})\widetilde{q}
-\sqrt{2}i(\overline{\xi}\widetilde{\psi})\sigma
\end{align}
for the 3d $\mathcal{N}=2$ chiral multiplets $Q$ and $\widetilde{Q}$ 
as well as the supersymmetry transformations 
(\ref{3dn2vmsusy1})-(\ref{3dn2vmsusy5}) and (\ref{3dn2cmadj1})-(\ref{3dn2cmadj3}), respectively, 
for the vector multiplet $V$ and the adjoint chiral multiplet $\Phi$. 
From the action (\ref{3dn4hm1c}) and the supersymmetric transformation laws, 
we obtain a  supercurrent of the 3d $\mathcal{N}=4$ hypermultiplets
\begin{align}
\begin{split}
J^{\mu}=
&-\sqrt{2}D^{\mu}\overline{q}\psi
-\overline{q} \gamma^{\mu} \overline{\lambda} q
+\sqrt{2}D_{\nu}\overline{q}\gamma^{\mu\nu}\psi
-\sqrt{2} \overline{q} \sigma \gamma^{\mu}\psi						\\
&-\sqrt{2} \widetilde{\psi} D^{\mu}\overline{\widetilde{q}}
+\widetilde{q}\gamma^{\mu}\overline{\lambda} \overline{\widetilde{q}}
+\sqrt{2} \gamma^{\mu\nu}\widetilde{\psi} D_{\nu}\overline{\widetilde{q}}
+\sqrt{2} \gamma^{\mu}\widetilde{\psi} \sigma\overline{\widetilde{q}}		\\
&-2\gamma^{\mu}(
\overline{q} \overline{\phi} \overline{\widetilde{\psi}}
+\overline{q} \overline{\psi}_{\phi} \overline{\widetilde{q}}
+\overline{\psi} \overline{\phi} \overline{\widetilde{q}})	\,	.
\end{split}
\end{align}
Using the 3d $\mathcal{N}=2$ notation, 
we get the supersymmetric boundary conditions for the 3d $\mathcal{N}=4$ hypermultiplets
\begin{align}
\begin{split}
0=
&-\sqrt{2}D_{2}\overline{q}(\xi\psi)
-\sqrt{2} (\xi\widetilde{\psi}) D_{2}\overline{\widetilde{q}}
-\overline{q} (\xi\gamma^{2}\overline{\lambda}) q
+\widetilde{q}(\xi\gamma^{2}\overline{\lambda}) \overline{\widetilde{q}}			\\
&+\sqrt{2}D_{\nu}\overline{q}(\xi\gamma^{2\nu}\psi)
+\sqrt{2} (\xi\gamma^{2\nu}\widetilde{\psi}) D_{\nu}\overline{\widetilde{q}}
-\sqrt{2} \overline{q} \sigma (\xi\gamma^{2}\psi)
+\sqrt{2} (\xi\gamma^{2}\widetilde{\psi}) \sigma\overline{\widetilde{q}}			\\
&+2 \overline{q} \overline{\phi} (\xi\gamma^{2}\overline{\widetilde{\psi}})
+2\overline{q} (\xi\gamma^{2}\overline{\psi}_{\phi}) \overline{\widetilde{q}}
+2 (\xi\gamma^{2}\overline{\psi}) \overline{\phi} \overline{\widetilde{q}}			\\
&+\sqrt{2} (\overline{\xi}\overline{\psi}) D_{2}q
+\sqrt{2}D_{2}\widetilde{q}(\overline{\xi}\overline{\widetilde{\psi}})
+\overline{q} (\overline{\xi}\gamma^{2}\lambda) q
-\widetilde{q}(\overline{\xi}\gamma^{2}\lambda) \overline{\widetilde{q}}			\\
&-\sqrt{2} (\overline{\xi}\gamma^{2\nu}\overline{\psi}) D_{\nu}q
-\sqrt{2}D_{\nu}\widetilde{q}(\overline{\xi}\gamma^{2\nu}\overline{\widetilde{\psi}})
+\sqrt{2} (\overline{\xi}\gamma^{2}\overline{\psi}) \sigma q
-\sqrt{2} \widetilde{q} \sigma (\overline{\xi}\gamma^{2}\overline{\widetilde{\psi}})		\\
&-2 (\overline{\xi}\gamma^{2}\widetilde{\psi}) \phi q
-2 \widetilde{q}(\overline{\xi} \gamma^{2}\psi_{\phi}) q
-2\widetilde{q}\phi(\overline{\xi}\gamma^{2}\psi) 	\, 	.
\end{split} \label{susybc2a}
\end{align}

One can generalize the boundary conditions and their solutions by introducing additional boundary degrees of freedom. 
Also, it would be intriguing to explore space of the solutions for given information. 
We defer these to later work. 
In this paper, we focus on the investigation of basic half-BPS boundary conditions for the hypermultiplets.
As in the previous discussion on the vector multiplet, we examine the half-BPS boundary conditions of the A- and B-types for the pure hypermultiplets in this subsection, and discuss the coupled hypermultiplets in next subsection.

\subsubsection{A-type boundary condition}
\label{sechma}

We are interested in A-type boundary conditions for bosonic fields given by $\gamma^2 \xi = \overline{\xi}$ and fermionic boundary conditions,
\begin{align}
\label{hasusybc1}
\gamma^{2}\psi&=e^{2i\varphi}\overline{\psi}, &
\gamma^{2}\lambda&=e^{2i\theta}\overline{\lambda},&
\gamma^{2}\psi_{\phi}&= e^{2i\theta}\overline{\psi}_{\phi} 
\end{align}
where $\varphi,\theta\in \mathbb{R}$ are constant phase parameters. 
Here and in the following, we consider the case $e^{2i\varphi} = - e^{2i \theta}$, \textit{i.e.}
\begin{align}
\label{hasusybc2}
\theta - \varphi&=\frac{\pi}{2}+\pi\mathbb{Z}	\,	,
\end{align}
but for the A-type condition the case $e^{2i\varphi} = e^{2i \theta}$ provides equivalent results to the ones obtained from \eqref{hasusybc2}.

From (\ref{a1r1})-(\ref{a1r2}) 
the generic boundary conditions (\ref{susybc2a}) for the hypermultiplets become\footnote{
To see the general form of the supersymmetric boundary conditions 
of the coupled hypermultiplets, 
we obtained the condition (\ref{susybc2a0}) by using the component-wise projection condition (\ref{hasusybc1}) 
given fixed all the gauge and global symmetry indices. 
Given the data of preserved gauge and global symmetries at the boundary, 
a large family of the boundary conditions can be constructed from the results below by 
restoring the form of representation. 
}
\begin{align}
\begin{split}
\label{susybc2a0}
0=
e^{-i\varphi}
\Bigl[
&-\sqrt{2}(e^{i\varphi}D_{2} \cdot \overline{q}+e^{-i\varphi}D_{2} \cdot q)(\xi \psi)
-\sqrt{2}(e^{i\varphi}D_{2} \cdot \overline{\widetilde{q}}+e^{-i\varphi}D_{2} \cdot\widetilde{q})(\xi\widetilde{\psi})	\\
&+\sqrt{2}(e^{i\varphi}D_{m} \cdot \overline{q}-e^{-i\varphi}D_{m} \cdot q)(\xi \gamma^{2m}\psi)
+\sqrt{2}(e^{i\varphi}D_{m} \cdot \overline{\widetilde{q}}
-e^{-i\varphi}D_{m} \cdot \widetilde{q})(\xi \gamma^{2m}\widetilde{\psi})		\\	\
&-\sqrt{2}(e^{i\varphi}\sigma\cdot \overline{q}+e^{-i\varphi}\sigma\cdot q)(\xi\gamma^{2}\psi)
+\sqrt{2}(e^{i\varphi}\sigma\cdot \overline{\widetilde{q}}+e^{-i\varphi}\sigma\cdot \widetilde{q})
(\xi\gamma^{2}\widetilde{\psi})		\\		
&+2(e^{-i\varphi}\overline{\phi}\cdot \overline{q}+e^{-i\varphi}\phi\cdot q)(\xi\widetilde{\psi})
+2(e^{i\varphi}\overline{\phi}\cdot \overline{\widetilde{q}}
+e^{-i\varphi}\phi\cdot \widetilde{q})(\xi\psi)
\Bigr]		\\	
+
e^{-i\theta}
\Bigl[&
-(q\overline{q}e^{-i\theta}+q\overline{q}e^{i\theta})(\xi\lambda)
+(\widetilde{q}\overline{\widetilde{q}}e^{-i\theta}+\widetilde{q}\overline{\widetilde{q}}e^{i\theta})(\xi\lambda)
+2(\overline{q}\overline{\widetilde{q}}e^{i\theta}+q\widetilde{q}e^{-i\theta})
(\xi\overline{\psi}_{\phi})
\Bigr]
\end{split}
\end{align}
where the dot $\cdot$ indicates the gauge and global symmetry action on the hypermultiplets. 
Also, the generators for gauge group are implicit between the products of two scalars, 
\textit{e.g.} $\overline{q} T^a_{R} q$.

We would like to find the solutions to the half-BPS boundary conditions of pure hypermultiplet for $\varphi=0$ and $\varphi=\frac{\pi}{2}$.\footnote{The half-BPS boundary conditions preserving $\mathcal{N}=(2,2)$ for hypermultiplet were also obtained in \cite{Dimofte:2012pd}.}

\begin{itemize}[leftmargin=5mm]

\item[i)] $\gamma^2 \psi =  \overline{\psi}$ (when $\varphi=0$)

In the case with $\varphi=0$,
we find from (\ref{susybc2a0}) the following bosonic boundary conditions for the hypermultiplets 
\begin{align}
\label{hasusybc1a}
\partial_{2}(\operatorname{Re} q)&=0, 			&\hspace{-20mm}	 \partial_{2}(\operatorname{Re} \widetilde{q})&=0,\\
\label{hasusybc1b}
\partial_{m}(\operatorname{Im} q)&=0, 		&\hspace{-20mm}	\partial_{m}(\operatorname{Im} \widetilde{q})&=0.
\end{align}

\item[ii)] $\gamma^2 \psi =  -\overline{\psi}$ (when $\varphi=\frac{\pi}{2}$)

For the other A-type boundary conditions with the fermionic boundary conditions $\varphi=\frac{\pi}{2}$,
the bosonic boundary conditions for the hypermultiplets read
\begin{align}
\label{hasusybc2a}
\partial_{m}(\operatorname{Re} q)&=0,		&\hspace{-20mm}		\partial_{m}(\operatorname{Re} \widetilde{q})&=0,\\
\label{hasusybc2b}
\partial_{2}(\operatorname{Im} q)&=0,			&\hspace{-20mm}		\partial_{2}(\operatorname{Im} \widetilde{q})&=0.
\end{align}

\end{itemize}

\subsubsection*{Boundary degree of freedom for the A-type from the pure bulk hypermultiplet}

The A-type conditions provide decomposition of 
the 3d $\mathcal{N}=4$ hypermultiplets into 
the boundary supermultiplets in such a way that 
($\text{Re}q$, $\text{Re}\widetilde{q}$) 
fluctuate at the boundary and ($\text{Im}q$, $\text{Im}\widetilde{q}$) 
satisfy Dirichlet boundary conditions, 
or the other way around.
Each of them 
forms the 2d $\mathcal{N}=(2,2)$ chiral multiplets 
$\Phi^{(2,2)}_{I}$ and ${\Phi}^{(2,2)}_{II}$ 
whose lowest components are the complex scalar fields,
which consists of 
$(\text{Re}q, \text{Re}(\widetilde{q}))$ and $(\text{Im}q, \text{Im}(\widetilde{q}))$,
respectively;
\begin{align}
\label{adec2}
H_{\mathcal{N}=4}&\rightarrow 
(\Phi^{(2,2)}_I, {\Phi}^{(2,2)}_{II}).
\end{align}

The A-type boundary conditions $\gamma^{2}\psi=\overline{\psi}$ give both left-moving and right-moving fermions. 
We may denote the two-dimensional fermions by 
\begin{align}
\label{22susycm1}
\psi^{+}&:=\frac12\left(\psi+\overline{\psi}\right)
+\frac{i}{2}\left(\widetilde{\psi}+\overline{\widetilde{\psi}}\right),& 
\psi^{-}&:=\frac{1}{2i}\left(\psi-\overline{\psi}\right)+\frac12\left(\widetilde{\psi}-\overline{\widetilde{\psi}}_{\phi}\right),
\end{align}
and their complex conjugate; $\overline{\psi}^{+}=(\psi^{+})^{\dag}$, $\overline{\psi}^{-}=(\psi^{-})^{\dag}$. 
They would carry the R-charges as
\begin{align}
\label{22susycm2}
\begin{array}{c|c|c|c}
&SO(1,1)&U(1)_{A}&U(1)_V\\ \hline
\psi^{+}&+&+&-\\
\psi^{-}&-&-&-\\
\overline{\psi}^{+}&+&-&+\\
\overline{\psi}^{-}&-&+&+\\
\end{array}
\end{align}
We also denote a two-dimensional complex scalar fields by
$\varphi:=\operatorname{Re} q+i\operatorname{Re} \widetilde{q}$. 
The supersymmetry transformations of component fields $(\varphi, \psi^{+}, \psi^{-}, \overline{\psi}^{+}, \overline{\psi}^{-})$, 
which form the chiral multiplet $\Phi^{(2,2)}_{I}$, would be given by
\begin{align}
\label{22susycm3}
\delta \varphi&=\xi_{+}\psi_{-}-\xi_{-}\psi_{+}\\
\label{22susycm4}
\delta \psi_{+}&=2i\partial_{+}\varphi\overline{\xi}_{-}+F\xi_{+}, \\
\label{22susycm5}
\delta \psi_{-}&=-2i\partial_{-}\varphi\overline{\xi}_{+}+F\xi_{-}, \\
\label{22susycm6}
\delta F&=-2i\overline{\xi}_{+}\partial_{-}\psi_{+}-2i\overline{\xi}_{-}\partial_{+}\psi_{-}.
\end{align}
where $F$ is an auxiliary field and \eqref{22susy1a} is used.
One can similarly realize the supersymmetric transformation laws of the other chiral superfield $\Phi^{(2,2)}_{II}$.

\subsubsection{B-type boundary conditions}
\label{sechmb}
The B-type conditions are characterized by the chiral projection (\ref{bsusybc1}) on supersymmetric parameter. 
We can find the bosonic boundary conditions by considering the fermionic boundary conditions 
\begin{align}
\label{hbsusybc1}
\gamma^{2}\psi = e^{2i\varphi}\psi, 			\qquad
\gamma^{2}\lambda = e^{2i\theta}\lambda,	\qquad
\gamma^{2}\psi_{\phi} = e^{2i\theta}\psi_{\phi} 
\end{align}
with $(\varphi,\theta)=(0,\frac{\pi}{2})$ and $(\frac{\pi}{2},0)$. 
When $(\varphi,\theta)=(0,\frac{\pi}{2})$, 
\textit{i.e.} $\gamma^2 \psi = \psi$, $\gamma^2 \lambda = -\lambda$, and $\gamma^{2}\psi_{\phi} = -\psi_{\phi}$ 
by using the formulae (\ref{b1r}) for $\lambda$, $\psi_{\phi}$ and (\ref{b2r}) for $\psi$, $\widetilde{\psi}$, 
we obtain
\begin{align}
\begin{split}
0=&-\sqrt{2}(D_{2} \cdot \overline{q}+\sigma \cdot \overline{q})(\xi\psi)
-\sqrt{2}(D_{2} \cdot \overline{\widetilde{q}}-\sigma \cdot \overline{\widetilde{q}})(\xi\widetilde{\psi})	\\
&+2 \overline{\phi} \cdot \overline{q}(\xi\overline{\widetilde{\psi}})
+2\overline{\widetilde{q}} \cdot \overline{\phi}(\xi \overline{\psi}) \, .
\end{split}		\label{hbsusybc3b1}
\end{align}
Similarly, 
when $(\varphi,\theta)=(\frac{\pi}{2},0)$, \textit{i.e.} 
$\gamma^2 \psi = -\psi$, $\gamma^2 \lambda = \lambda$, and $\gamma^{2}\psi_{\phi} = \psi_{\phi}$, 
the boundary condition becomes
\begin{align}
\begin{split}
0=
&\sqrt{2}D_{m} \cdot \overline{q}(\xi\gamma^{m}\psi)
+\sqrt{2}D_{m} \cdot \overline{\widetilde{q}}(\xi\gamma^{m}\widetilde{\psi})	\\
&-\left(|q|^{2}-|\widetilde{q}|^{2}\right)(\xi\overline{\lambda})
+2\overline{q}\overline{\widetilde{q}}(\xi\overline{\psi}_{\phi}) \, .
\end{split}		\label{hbsusybc3a1}
\end{align}
The chiralities of the fermionic fields at the boundary from the bulk 3d $\mathcal{N}=4$ hypermultiplet are determined by the phase factor $\varphi\in \mathbb{R}$. 
For $\varphi=0$ (resp. $\varphi=\frac{\pi}{2}$), 
the right-moving (resp. left-moving) 
fermions survive at the two-dimensional boundary.

\begin{itemize}[leftmargin=5mm]

\item[i)] $\gamma^{2}\psi = \psi$ (when $\varphi=0$)

This boundary conditions admit the right-moving fermions with $\varphi=0$ in the hypermultiplets. 
For pure hypermultiplet, we turn off fields from the vector multiplet, 
so the condition (\ref{hbsusybc3b1}) leads to the Neumann boundary conditions for the hypermultiplet scalars $q$ and $\widetilde{q}$
\begin{align}
\label{hbsusybc3b2}
\partial_{2}q = 0 \, ,		\qquad		\partial_{2} \widetilde{q} = 0 \, .
\end{align}

\item[ii)] $\gamma^{2}\psi = -\psi$ (when $\varphi=\frac{\pi}{2}$)

In this case, the fermions in hypermultiplet at the boundary are the left-moving. 
For the hypermultiplets without gauge coupling, 
(\ref{hbsusybc3a1}) can be solved by requiring the Dirichlet boundary conditions 
for the hypermultiplet scalars $q$ and $\widetilde{q}$
\begin{align}
\label{hbsusybc3a2}
\partial_{m}q = 0 \, , 		\qquad		\partial_{m}\widetilde{q} = 0 \, .
\end{align}
Therefore the bosonic degrees of freedom in the 3d $\mathcal{N}=4$ hypermultiplets cannot survive at the boundary 
while the left-moving fermions are free to fluctuate at the boundary.

\end{itemize}

\subsubsection*{Boundary degree of freedom for the B-type from the pure bulk hypermultiplet}

We saw that there are two types of the B-type conditions 
for the 3d $\mathcal{N}=4$ hypermultiplets. 
For the boundary condition i) with $\varphi = 0$, 
the full set of four bosonic fields as well as the right-moving fermions $(\bm{2},\bm{1})_{+}$ can fluctuate at the boundary. 
They are packaged into the 2d $\mathcal{N}=(0,4)$ hypermultiplets $H^{(0,4)}$.
On the other hand, for the second conditions ii) with $\varphi = \frac{\pi}{2}$,
the left-moving fermions $(\bm{1},\bm{1})_{-}$ can fluctuate but 
all the bosonic degrees of freedom satisfy 
Dirichlet condition at the boundary.
The fluctuating 
degrees of freedom 
can be packaged into the $\mathcal{N}=(0,4)$ Fermi multiplets 
$\Lambda^{(0,4)}$. 
Therefore, we have the decomposition 
\begin{align}
\label{bdec2}
H_{\mathcal{N}=4}&\rightarrow 
(H^{(0,4)}, \Lambda^{(0,4)}).
\end{align}

The B-type boundary conditions $\gamma^{2}\psi=\psi$ lead to the right-moving fermions. 
We write them as $\Psi^{+AY}$ where the indices $Y=1,2$ represent the doublet under the additional $SU(2)_F$ global symmetry, 
\begin{align}
\label{04susyhm1}
\Psi^{+11}		:=	\overline{\widetilde{\psi}}	\,	,	\qquad
\Psi^{+12}		:=	\overline{\psi}	\,	,	\qquad
\Psi^{+21}		:=	\psi	\,	,	\qquad
\Psi^{+22}		:=	-\widetilde{\psi}	\,	.
\end{align}
Also, we denote the scalar component by
\begin{align}
\label{04susyhm2}
X^{\dot{1}1}	:=	q	\,	,	\qquad
X^{\dot{1}2}	:=	-\widetilde{q}	\,	,	 \qquad
X^{\dot{2}1}	:=	\overline{\widetilde{q}}	\,	,	\qquad
X^{\dot{2}2}	:=	\overline{q}	\,	,
\end{align}
which transform as a doublet under the $SU(2)_H$ and a doublet under the $SU(2)_F$.
The supersymmetry transformation of component fields $(X^{\dot{A}Y}, \Psi^{+AY})$, which forms a hypermultiplet $H^{(0,4)}$, can be obtained from 3d $\mathcal{N}=4$ supersymmetry transformation with projection and boundary conditions,
\begin{align}
\label{04susyhm3}
\delta X^{\dot{A}Y}	&=	-\sqrt{2} \xi^{-B\dot{A}} \epsilon_{BC} \Psi^{+CY}	\,	,	\\
\label{04susyhm4}
\delta \Psi^{+AY}		&=	\xi^{+A\dot{B}} \epsilon_{\dot{B}\dot{C}} (\partial_0 + \partial_1) X^{\dot{C}Y}	\,	, 
\end{align}
which is a supersymmetry transformation of the $\mathcal{N}=(0,4)$ hypermultiplet.

Another B-type boundary condition $\gamma^{2}\psi=-\psi$ leads to four real left-moving fermions, which are singlet under the R-symmetry. These fermionic fields form a Fermi multiplet $\Lambda^{(0,4)}$.
We can take them as two complex fermions, which we denote as
\begin{align}
\zeta^{-}_{1} = \psi		\,	,	\qquad	\zeta^{-}_{2} = \overline{\widetilde{\psi}}
\end{align}
where hermitian conjugates are $\overline{\zeta}^{-}_{1} = \overline{\psi}$ and $\overline{\zeta}^{-}_{2} = \widetilde{\psi}$, respectively.
Then, the supersymmetry transformation of these fields can be obtained and they are
\begin{align}
\delta \zeta^{-}_{1}	&=	-\sqrt{2} i \xi^{-A \dot{A}} \epsilon_{AB} \epsilon_{\dot{A}\dot{B}} \widetilde{X}^{B1'} X^{\dot{B}}	\\
\delta \zeta^{-}_{2}	&=	-\sqrt{2} i \xi^{-A \dot{A}} \epsilon_{AB} \epsilon_{\dot{A}\dot{B}} \widetilde{X}^{B2'} X^{\dot{B}}	\,	.	
\end{align}
These can be organized into
\begin{align}
\delta \Theta^{-Y' Y} = -\sqrt{2} i \xi^{-A \dot{A}} \epsilon_{AB} \epsilon_{\dot{A}\dot{B}} \widetilde{X}^{BY'} X^{\dot{B}Y}
\end{align}
where $\Theta^{Y' Y} = \begin{pmatrix} \zeta^{-}_{1} & \overline{\zeta}^{-}_{2} \\ \zeta^{-}_{2} & -\overline{\zeta}^{-}_{1} \end{pmatrix}$, which is a supersymmetry transformation of the Fermi multiplet.
When considering generic interaction with boundary fields, the supersymmetry transformation would take a form
\begin{align}
\label{04susyfm2}
\delta \zeta^{-}_{a}	=	-\sqrt{2}	i	\xi^{-}_{A\dot{A}} C_{a}^{A\dot{A}}
\end{align}
where $\zeta^{-}_{a}, a=1,2,3,4$ denotes $\zeta^{-}_{1}, \zeta^{-}_{2}, \overline{\zeta}^{-}_{1}$, and $\overline{\zeta}^{-}_{1}$ and $C^{A\dot{A}}_{a}$ are some function of $X^{\dot{A}Y}$ and $\widetilde{X}^{AY'}$. See also \cite{Witten:1994tz, Gates:1994bu}.

\subsection{Gauge coupling and SUSY deformations}
\label{secgeneral}

We now discuss the half-BPS boundary conditions 
for the vector multiplets and the hypermultiplets in the 3d $\mathcal{N}=4$ supersymmetric gauge theories
with the supersymmetric deformation by FI parameters and mass parameters.

\subsubsection{FI and mass deformations}
\label{secdef}

The 3d $\mathcal{N}=4$ gauge theories can be deformed 
by Fayet-Iliopoulos (FI) terms and mass terms
while keeping supersymmetry. 
We consider the effects of the supersymmetric deformations on the half-BPS boundary conditions.

If the 3d $\mathcal{N}=4$ supersymmetric gauge theories involve the $U(1)$ factors of the gauge group, 
they can be deformed in a supersymmetric way by introducing the BF coupling of  
the topological currents for the $U(1)$ factors to a background Abelian $\mathcal{N}=4$ twisted vector multiplet 
$(V_{r},\Phi_{r})$ \cite{Brooks:1994nn,Kapustin:1999ha}
\begin{align}
\begin{split}
\label{fidef1}
S_{FI}
&=\int d^{3}x d^{4}\theta\ 
\mathrm{Tr}' 
\left(\Sigma V_{r}\right)	\\	
&+ \frac{i}{2} \int d^{3}x d^{2}\theta\ 
\mathrm{Tr}'
\left(\Phi\Phi_{r}\right) + c.c.
\end{split}
\end{align}
where $V_{r}=ir \bar{\theta}\theta$ and $\Phi_{r}=\phi_{r}$ 
with $r\in \mathbb{R}$, $\phi_{r}\in \mathbb{C}$.  
The trace $\mathrm{Tr}'$ only takes the $U(1)$ factors of the gauge group.
Here 
$r^{\hat{i}} = (r,\operatorname{Re}(\phi_{r}),\operatorname{Im}(\phi_{r}))$, $\hat{i}=7,8,9$, 
forms a triplet under the $SU(2)_{H}$. 
In terms of the component fields, 
we can express the action (\ref{fidef1}) as
\begin{align}
\label{fidef2}
S_{FI}=&\int d^{3}x\ 
\left[
-\frac{1}{2} rD + \frac{i}{2} \phi_{r}F_{\phi} - \frac{i}{2} \overline{\phi}_{r}\overline{F}_{\phi} 
\right]
\end{align}
where $r$ is a real FI parameter 
and $\phi_{r}$ a complex FI parameter. 
The conserved supercurrent is 
\begin{align}
\label{fidefcur1}
J_{r}^{\mu}& = \frac{1}{2}r(\gamma^{\mu}\overline{\lambda}) 
+\frac{\sqrt{2}}{2} \overline{\phi}_{r}(\gamma^{\mu}\overline{\psi}_{\phi}).
\end{align}

One can also deform the 3d $\mathcal{N}=4$ supersymmetric gauge theories 
in a supersymmetric way by introducing 
mass terms for the hypermultiplets. 
It can be achieved by coupling $Q$ and $\widetilde{Q}$ 
to a background Abelian $\mathcal{N}=4$ vector multiplet 
$(V_{M},\Phi_{M})$ 
\begin{align}
\label{massdef1}
\begin{split}
S_{M}=&-\int d^{3}x d^{4}\theta 
\left(
\overline{Q} e^{-2V_{M}}Q+\overline{\widetilde{Q}}e^{2V_{M}}\widetilde{Q}
\right) \\
&+\sqrt{2}i\int d^{3}x d^{2}\theta 
\left(
\widetilde{Q}\Phi_{M}Q
\right)+c.c.
\end{split}
\end{align}
where $V_{M}=i{M}\bar{\theta}\theta$ and $\Phi_{M}=\phi_{M}$ 
and ${M}\in \mathbb{R}$ is real mass and $\phi_{M}\in \mathbb{C}$ is complex mass parameters.  
Here $({M}, \operatorname{Re}(\phi_{M}),\operatorname{Im}(\phi_{M}))$ 
forms a triplet under the $SU(2)_{C}$. 
In the component fields, 
the action (\ref{massdef1}) can be expressed as 
\begin{align}
\label{massdef2}
\begin{split}
S_{M}=&\int d^{3}x\ 
\Biggl[
-{M}^{2}\left(|q|^{2}+|\widetilde{q}|^{2} \right)
-i{M}\left(\overline{\psi}\psi-\overline{\widetilde{\psi}}\widetilde{\psi} \right)
-(2F_{q}\overline{F}_{q}-2F_{\widetilde{q}}\overline{F}_{\widetilde{q}})	\\
&
+\sqrt{2}i\phi_{M}\left(
F_{\widetilde{q}}q+F_{q}\widetilde{q}
\right)
-i \phi_{M}\widetilde{\psi}\psi
+\sqrt{2}i \overline{\phi}_{M}\left(
\overline{F}_{\widetilde{q}}\overline{q}+
\overline{F}_{q}\overline{\widetilde{q}}
\right)
+i \overline{\phi}_{M}
\overline{\widetilde{\psi}}\overline{\psi}
\Biggr] \; .
\end{split}
\end{align}
The conserved supercurrent is 
\begin{align}
\label{masscur1}
\begin{split}
J_{M}^{\mu}
=&-\sqrt{2}{M} \overline{q}\gamma^{\mu}\psi
+\sqrt{2}{M} \overline{\widetilde{q}}\gamma^{\mu}\widetilde{\psi}		\\
&
+2\overline{\phi}_{M}\overline{q}\gamma^{\mu}\overline{\widetilde{\psi}}
+2\overline{\phi}_{M}\overline{\widetilde{q}}\gamma^{\mu}\overline{\psi}.
\end{split}
\end{align}

The supercurrents (\ref{fidefcur1}) and (\ref{masscur1}) provide 
additional contributions to the supercurrents we obtained in previous sections
and modify the supersymmetric boundary conditions.

\subsubsection{Coupled Hypermultiplets}
\label{secgenehm}
We consider the half-BPS boundary conditions for the coupled hypermultiplet with FI and mass 
parameters turned on. 
Due to the coupling, the half-BPS boundary conditions for the hypermultiplets
depends on the choice of the half-BPS boundary conditions for the vector multiplet 
discussed in subsection \ref{secvm1} with condition \eqref{hasusybc2}.
This provides a large class of the half-BPS boundary conditions 
specified by the preserved gauge and flavor symmetries at the boundary. 
Here we want to find general structure of deformed boundary conditions 
for the hypermultiplets due to gauge coupling, FI parameters, and mass parameters.

\subsubsection*{$\bullet$ \textnormal{\textit{A-type boundary conditions}}}

\begin{itemize}[leftmargin=5mm]

\item[i)] $\gamma^2 \psi = \overline{\psi}$, $\gamma^2 \lambda = -\overline{\lambda}$, and $\gamma^2 \psi_\phi = -\overline{\psi}_\phi$ 
(when $\varphi=0$, $\theta=\frac{\pi}{2}$)

For the A-type conditions with 
$(\varphi,\theta)=(0,\frac{\pi}{2})$,  
we find from \eqref{susybc2a0} the generic half of supersymmetric boundary conditions for the hypermultiplets 
\begin{align}
\label{dhasusybc1a}
&D_{2} \cdot (\operatorname{Re} q)=
\sqrt{2}\operatorname{Re}
\left[(\phi+\phi_{M})\cdot \widetilde{q}\right],& 
&D_{2} \cdot (\operatorname{Re} \widetilde{q})=
\sqrt{2}\operatorname{Re}
\left[(\phi+\phi_{M})\cdot q\right],\\
\label{dhasusybc1b}
&D_{m} \cdot (\operatorname{Im} q)=0,& 
&D_{m} \cdot (\operatorname{Im} \widetilde{q})=0,\\
\label{dhasusybc1c0}
&(\sigma+{M})\cdot (\operatorname{Re} q)=0,& 
&(\sigma+{M})\cdot (\operatorname{Re} \widetilde{q})=0,\\
\label{dhasusybc2d}
&
\operatorname{Im} (\widetilde{q}q)
=\operatorname{Im} (\phi_{r}).
\end{align}

The conditions (\ref{dhasusybc1a}) say that 
the real parts of the complex scalar fields 
$q$ and $\widetilde{q}$ can fluctuate 
while satisfying the Robin-type boundary conditions, 
which specify a linear combination of the fields 
and the normal components of their derivatives at the boundary. 
The conditions (\ref{dhasusybc1b}) imply 
that the imaginary parts of $q$ and $\widetilde{q}$ are subject to the Dirichlet-like boundary conditions.

The other set 
(\ref{dhasusybc1c0}) and (\ref{dhasusybc2d})
are algebraic constraints which are responsible for the gauge coupling. 
The precise forms of the boundary conditions and the possible solutions 
depend on the detail of the 3d $\mathcal{N}=4$ vector multiplet and hypermultiplets 
but these equations can be regarded as the basic building blocks of boundary conditions.

The real parts of $q$ and $\widetilde{q}$, 
which are fluctuating degrees of freedom at the boundary, satisfy conditions (\ref{dhasusybc1c0}).
As a coupled vector multiplet satisfies the magnetic-like A-type boundary conditions when $\theta=\pi/2$, 
the two vector multiplet scalars $\operatorname{Re} \phi$ and $\sigma$ 
obey the Dirichlet boundary conditions (\ref{asusybc3b}). 
Thus the constraints (\ref{dhasusybc1c0}) can be solved by setting $\sigma$ to specific fixed values at the boundary.

The last condition \eqref{dhasusybc2d} doesn't involve any bosonic fields in the vector multiplet, 
but it appears due to the gauge coupling and FI deformations as it is induced from the fermionic bilinear form involving $\psi_{\phi}$. 
It is an imaginary part of the complex moment map $\mu_{\mathbb{C}}$ with fields restricted at the boundary.

\item[ii)] $\gamma^2 \psi = -\overline{\psi}$, $\gamma^2 \lambda = \overline{\lambda}$, and $\gamma^2 \psi_\phi = \overline{\psi}_\phi$ 
(when $\varphi=\frac{\pi}{2}$, $\theta=0$)

The A-type conditions with $(\varphi,\theta)=(\frac{\pi}{2},0)$ are 
\begin{align}
\label{dhasusybc2c}
&D_{2} \cdot (\operatorname{Im} q) = \sqrt{2}\operatorname{Im} \Bigl[(\phi+\phi_{M})\cdot \widetilde{q}\Bigr],	& 
&D_{2} \cdot (\operatorname{Im} \widetilde{q}) = \sqrt{2}\operatorname{Im} \Bigl[(\phi+\phi_{M})\cdot q\Bigr],	\\	
\label{dhasusybc2b}
&D_{m} \cdot (\operatorname{Re} q)=0,		& 
&D_{m} \cdot (\operatorname{Re} \widetilde{q})=0,\\	
\label{dhasusybc2c0}
&(\sigma+{M})\cdot (\operatorname{Im} q)=0,& 
&(\sigma+{M})\cdot (\operatorname{Im} \widetilde{q})=0,
\\
\label{dhasusybc1d}
&
\operatorname{Re} (\widetilde{q} q)=
\operatorname{Re} (\phi_{r}),&  
&
|q|^{2}-|\widetilde{q}|^{2}=r.
\end{align}

In this case, the real parts of $q$ and $\widetilde{q}$ satisfy Dirchlet-like boundary conditions \eqref{dhasusybc2b} and
the imaginary parts of $q$ and $\widetilde{q}$ can fluctuate while satisfying the Robin-type boundary conditions \eqref{dhasusybc2c}.
Again the remaining algebraic constraints (\ref{dhasusybc2c0}) and (\ref{dhasusybc1d}) 
arise from the coupling of the 3d $\mathcal{N}=4$ hypermultiplets to the vector multiplet.

For $\theta=0$, 
the vector multiplet is subject to the electric A-type boundary conditions,  
where only the vector multiplet scalar $\operatorname{Im} \phi$ satisfies the Dirichlet-like boundary condition 
(\ref{asusybc3d0}) and other scalars satisfy the Neumann-like boundary condition \eqref{asusybc3d}.
As $\sigma$ can flucuate at the boundary, the constraint (\ref{dhasusybc2c0}) are the conditions for the coupling of the hypermultiplet scalar fields $q$, $\widetilde{q}$ 
and the vector multiplet scalar $\sigma$ in a supersymmetric way when considering a boundary superpotential.

Two conditions in (\ref{dhasusybc1d}) are the constraints on the bulk hypermultiplets 
due to the gauge coupling and FI deformations, 
which are a real part of the complex moment map $\mu_{\mathbb{C}}$ and the real moment map $\mu_{\mathbb{R}}$, respectively, 
with fields restricted at the boundary.

\end{itemize}

\subsubsection*{$\bullet$ \textnormal{\textit{B-type boundary conditions}}}

\begin{itemize}[leftmargin=5mm]

\item[i)] $\gamma^2 \psi = \psi$, $\gamma^2 \lambda = -\lambda$, and $\gamma^2 \psi_\phi = -\psi_\phi$ (when $\varphi=0$ and $\theta=\frac{\pi}{2}$)

As the full R-symmetry $SU(2)_{C}\times SU(2)_{H}$ is maintained for the B-type conditions, 
a pair of fermionic fields $\psi$ and $\overline{\widetilde{\psi}}$ may form the supermultiplet. 
For $(\varphi,\theta)=(0,\frac{\pi}{2})$, 
we obtain, from (\ref{hbsusybc3b1}), (\ref{fidefcur1}) and (\ref{masscur1}), 
\begin{align}
\label{dhbsusybc3b2}
D_{2} \cdot q + (\phi^{i}+{M}^{i})\cdot q&=0,		&		D_{2} \cdot \widetilde{q}-(\phi^{i}+{M}^{i})\cdot \widetilde{q}&=0
\end{align}
where we have defined the triplet 
$\phi^{i}=(\sigma, \operatorname{Re}\phi,\operatorname{Im}\phi)$
and 
${M}^{i}=({M}, \operatorname{Re}\phi_{M},\operatorname{Im}\phi_{M})$
of the $SU(2)_{C}$. 
The bosonic degrees of freedom $q$ and $\widetilde{q}$ can fluctuate at the boundary 
while satisfying Robin-type boundary conditions \eqref{dhbsusybc3b2}. 
In this case, the vector multiplet obeys 
the electric B-type conditions that admit the Dirichlet-like boundary condition (\ref{bsusybc3a}) for all the vector multiplet scalars. 
Also, 
the detail forms of boundary conditions depend on the specific data of the theories, however, (\ref{dhbsusybc3b2}) can be viewed as the 
basic building blocks for the boundary conditions.

\item[ii)] $\gamma^2 \psi = -\psi$, $\gamma^2 \lambda = \lambda$, and $\gamma^2 \psi_\phi = \psi_\phi$ (when $\varphi= \frac{\pi}{2}$ and $\theta=0$)

From (\ref{hbsusybc3a1}), (\ref{fidefcur1}) and (\ref{masscur1}),
the general B-type conditions with 
$(\varphi,\theta)=(\frac{\pi}{2},0)$ read 
\begin{align}
\label{dhbsusybc3a2}
D_{m} \cdot q&=0,& D_{m} \cdot \widetilde{q}&=0,
\\ \label{dhbsusybc3a3}
|q|^{2}-|\widetilde{q}|^{2}&=r,& q\widetilde{q}&=\phi_{r}.
\end{align}

Similarly as before, the algebraic conditions (\ref{dhbsusybc3a3}) come from 
the gauge coupling and FI deformation.

\end{itemize}

\subsubsection{BPS boundary conditions and 3d $\mathcal{N}=4$ vacua}
\label{secbpsvacua}

The classical moduli space of the 3d $\mathcal{N}=4$ supersymmetric gauge theory on $\mathbb{R}^{1,2}$ 
is determined by the set of equations 
\begin{align}
\label{modsp1a1}
[\phi^{i},\phi^{j}]&=0,\\
\label{modsp1a2}
(\phi^{i}+{M}^{i})\cdot \left(q,\widetilde{q}\right)&=0,\\
\label{modsp1a3}
\mu^{\hat{i}}+r^{\hat{i}}&=0
\end{align}
where the dot $\cdot$ implies the action of the 
gauge and flavor symmetry group on the hypermultiplet scalars $(q,\widetilde{q})$. 
Here $\mu^{\hat{i}}$ are the three hyperk\"{a}hler moment maps 
for the action of the gauge symmetry group on the hypermultiplets. 
They split into the real and complex moment maps 
$\mu_{\mathbb{R}}$ and $\mu_{\mathbb{C}}$ \cite{Bullimore:2016nji}. 
They are respectively associated to the K\"{a}hler form 
\begin{align}
\label{modsp1b}
\omega&=\sum_{I}\left(
dq^{I}\wedge d\overline{q}^{I}
+d\widetilde{q}^{I}\wedge d\overline{\widetilde{q}}^{I}
\right)
\end{align}
and the holomorphic symplectic form 
\begin{align}
\label{modsp1c}
\Omega&=\sum_{I}
\left(
dq^{I}\wedge d\widetilde{q}^{I}
\right)
\end{align}
and given by 
\begin{align}
\label{modsp1d1}
\mu_{\mathbb{R}}&=
|q|^{2}-|\widetilde{q}|^{2},\\
\label{modsp1d2}
\mu_{\mathbb{C}}&=
q\widetilde{q}.
\end{align}

We remark that 
the half-BPS boundary conditions detect 
the set of the defining equations (\ref{modsp1a1})-(\ref{modsp1a3}) 
of the vacua. 
We have encountered the equation (\ref{modsp1a1}) 
in the vector multiplet boundary conditions 
(\ref{asusybc3b0}) and (\ref{bsusybc3d}) where fields are restricted at the boundary, 
which can be expected as it characterizes the Coulomb branch. 
The second set of equations (\ref{modsp1a2}) 
specify the coupling between the vector multiplet scalars 
and the hypermultiplet scalars. 
We have met these equations with fields restricted on the boundary
in the boundary conditions constraining the fluctuation of hypermultiplet scalars. 
As (\ref{modsp1a2}) suggests, 
these conditions can be shifted by turning on the mass parameters 
$\sigma \rightarrow \sigma+{M}, \phi \rightarrow \phi+\phi_{M}$.
We also saw that the moment maps \eqref{modsp1a3} with FI parameters 
appear as algebraic constraints for the scalar component of hypermultiplets at the boundary.

\section{Brane construction}
\label{secbrane}

In this section, we propose the brane configurations in Type IIB string theory 
corresponding to the half-BPS boundary conditions of the 3d $\mathcal{N}=4$ supersymmetric theories
discussed in section \ref{sec3dn4a}. 
We also study map between boundary supermultiplets arising from 3d bulk supermultiplets for simplest examples 
by considering $S$-duality of Type IIB theory.

\subsection{Type IIB configuration}
\label{sectypeiib}

We consider the brane realization of 3d $\mathcal{N}=4$ theories in Type IIB string theory 
on $\mathbb{R}^{1,9}$ \cite{Hanany:1996ie}.
Let $Q_{L}$ (resp. $Q_{R}$) be the supercharge 
generated by the left- (resp. right-) moving 
world-sheet degrees of freedom which 
satisfies the chirality condition of Type IIB string theory 
\begin{align}
\label{iib1}
\Gamma_{0123456789}Q_{L}=Q_{L} \, ,		\qquad		\Gamma_{0123456789}Q_{R}=Q_{R} \, .
\end{align} 
We consider D3-branes supported on $(x^{0},x^{1},x^{2},x^{6})$ and bounded along $x^6$ direction by two NS5-branes supported on $(x^{0},x^{1},x^{2},x^{3},x^{4},x^{5})$ or by two D5-branes supported on $(x^{0},x^{1},x^{2},x^{7},x^{8},x^{9})$
\begin{align}
\label{d3nsd5}
\begin{array}{ccccccccccc}
&0&1&2&3&4&5&6&7&8&9\\
\textrm{D3}
&\circ&\circ&\circ&-&-&-&\circ&-&-&- \\
\textrm{NS5}
&\circ&\circ&\circ&\circ&\circ&\circ&-&-&-&- \\
\textrm{D5}
&\circ&\circ&\circ&-&-&-&-&\circ&\circ&\circ \\
\end{array}
\end{align}
where $\circ$ denotes the directions in which branes are supported 
whereas $-$ stands for the directions at which branes are located.
The brane configuration (\ref{d3nsd5}) preserves 
linear combination of supercharges
$\epsilon_{L}Q_{L}+\epsilon_{R}Q_{R}$  
with
\begin{align}
\label{ns5p1}
\Gamma_{012345}\epsilon_{L}=\epsilon_{L} \, ,		\qquad		\Gamma_{012345}\epsilon_{R} =-\epsilon_{R}
\end{align}
and
\begin{align}
\label{d5p1}
\Gamma_{012789}\epsilon_{R}&=\epsilon_{L}	\,	,	\\
\label{d3p1}
\Gamma_{0126}\epsilon_{R}&=\epsilon_{L}	\,	.
\end{align}
Here, the first condition (\ref{ns5p1}) is the projection condition 
on spinors $\epsilon_{L}$ and $\epsilon_{R}$ imposed by the NS5-branes 
while (\ref{d5p1}) and (\ref{d3p1}) 
are the conditions by the the D5-branes and the D3-branes, respectively. 
From (\ref{ns5p1})-(\ref{d3p1}), 
we can find two non-trivial conditions on the spinors. 
So there remain 8 supercharges.

As D3-branes are bounded in $x^6$-direction, 
the low-energy effective theory of worldvolume of D3 branes are described 
by 3d $\mathcal{N}=4$ supersymmetric theories after decoupling the gravity.
The above brane configuration breaks the Lorentz symmetry group $SO(1,9)$ 
into $SO(1,2)_{012}\times SO(3)_{345}\times SO(3)_{789}$ 
where $SO(1,2)$ is Lorentz symmetry and the double covers of 
$SO(3)_{345} \times SO(3)_{789}$ give $SU(2)_{C}\times SU(2)_{H} \cong SO(4)_R$ 
R-symmetry of 3d $\mathcal{N}=4$ theories.

\subsection{D3-NS5 branes}
\label{secd3nsns}

Let us first consider the case 
where the $N$ coincident D3-branes are stretched 
between the two parallel NS5-branes. 
The low-energy effective theory is the 3d $\mathcal{N}=4$ $U(N)$ pure SYM theory \cite{Hanany:1996ie}. 
The three-dimensional coupling constant $g_{3d}^{2}$ is classically given by 
$\frac{1}{g_{3d}^{2}}
=\frac{\Delta x^{6}(\textrm{NS5})}{g_{4d}^{2}}
$
where $\Delta x^{6}(\textrm{NS5})$ is the interval of the stretched D3-branes along $x^{6}$ 
and $g_{4d}^{2}$ is the gauge coupling of 4d $\mathcal{N}=4$ SYM theory. 
The bosonic massless modes of the worldvolume theory of D3-branes are 
the fluctuations of the D3-branes in transverse directions $x^{3},x^{4},x^{5}$ and three-dimensional gauge fields. 
The $U(N)$ gauge symmetry has a non-trivial center $U(1)$, 
which parametrizes the motion of the center of mass of the $N$ D3-branes. 
The FI parameters $\{ r, \text{Re}(\phi_r), \text{Im}(\phi_r) \}$ are described by 
the relative positions of two NS5-branes along $x^{7}$, $x^{8}$, and $x^{9}$.

\subsubsection{A-type boundary conditions}
\label{secd3nsnsa}
\begin{figure}[h]
\centering
    \begin{tabular}{cc}
          \includegraphics[scale=0.5]{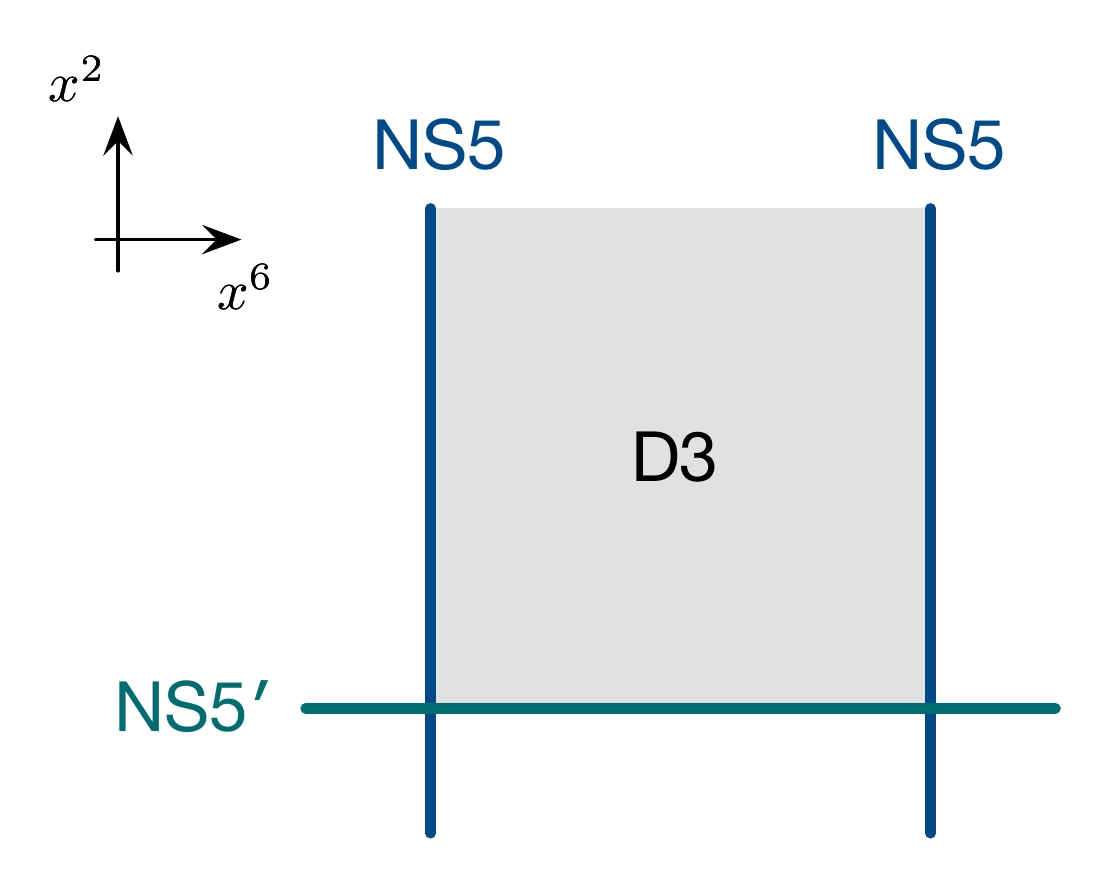}&\includegraphics[scale=0.5]{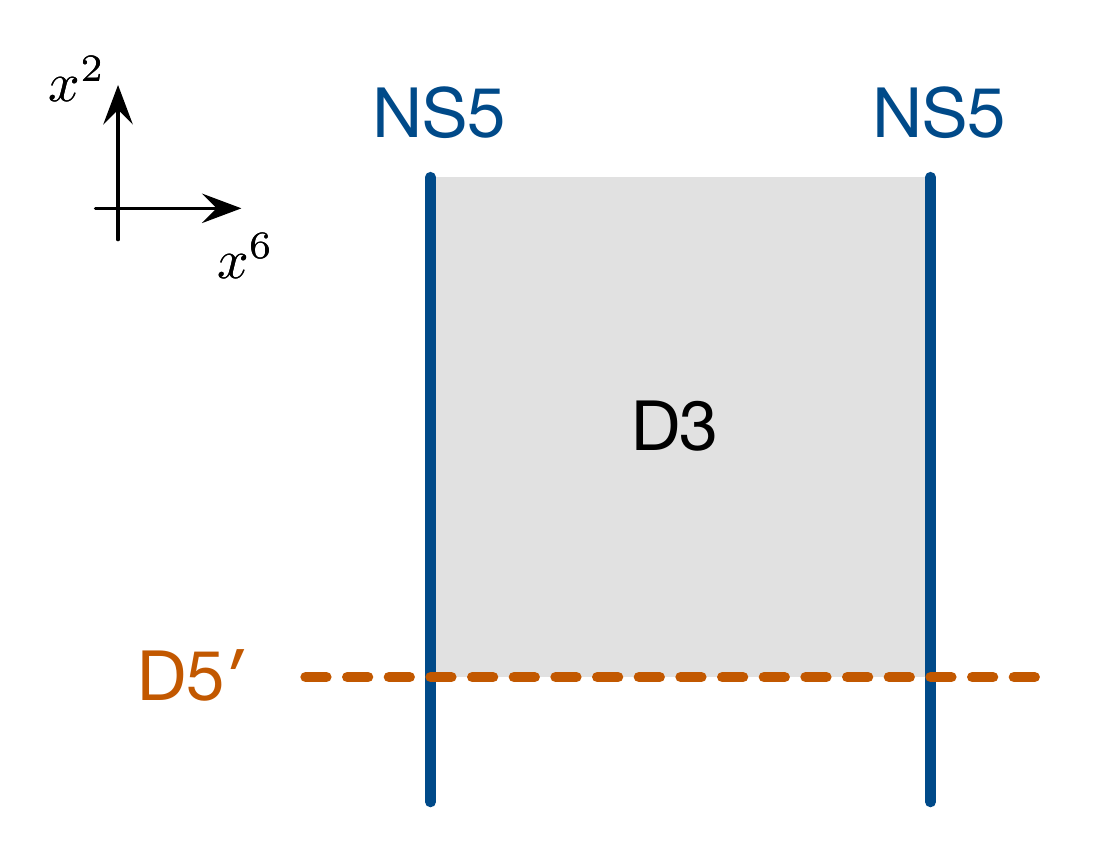}\\
          (a)NS5$'$-brane&(b)D5$'$-brane
\end{tabular}
    \caption{D3-NS5 system with NS5$'$-brane or D5$'$-brane. 
    NS5$'$-(D5$'$-)brane provides the electric-(magnetic-)like A-type boundary conditions for the vector multiplet where 2d $\mathcal{N}=(2,2)$ vector multiplet (twisted chiral multiplet) can fluctuate at the boundary.
    } 
    \label{fig10}
\end{figure}

The half-BPS boundary conditions for the pure 3d $\mathcal{N}=4$ vector multiplet discussed in subsection \ref{secvm1} 
can be realized in 
D3-NS5 brane system by introducing additional branes.
We call such additional branes NS5$'$-brane and D5$'$-brane 
where they are supported on $(x^{0},x^{1},x^{3},x^{4},x^{6},x^{9})$ 
and $(x^{0},x^{1},x^{5},x^{6},x^{7},x^{8})$, respectively. They are located at $x^{2}=0$ 
and D3-branes are extended in the half space $x^{2} \geq 0$.
\begin{align}
\label{nsnsbc1}
\begin{array}{ccccccccccc}
&0&1&2&3&4&5&6&7&8&9\\
\textrm{D3}
&\circ&\circ&\circ&-&-&-&\circ&-&-&- \\
\textrm{NS5}
&\circ&\circ&\circ&\circ&\circ&\circ&-&-&-&- \\
\textrm{NS5$'$}
&\circ&\circ&-&\circ&\circ&-&\circ&-&-&\circ \\
\textrm{D5$'$}
&\circ&\circ&-&-&-&\circ&\circ&\circ&\circ&-\\
\end{array}
\end{align}
Therefore the additional 5-branes provide 
the two-dimensional boundary at $x^2=0$ in the effective 3d $\mathcal{N}=4$ SYM theories (see Figure \ref{fig10}). 
Also, the original $SO(1,2) \times SO(3)_{345} \times SO(3)_{789}$ symmetry is broken to $SO(1,1) \times SO(2)_{34} \times SO(2)_{78}$.

The NS5$'$-brane and D5$'$-brane provide additional projection conditions, respectively,
\begin{align}
\begin{split}
\Gamma_{013469}\epsilon_{L}=\epsilon_{L}, \quad &
\Gamma_{013469}\epsilon_{R}=-\epsilon_{R}	\, ,
\end{split} \label{nsns1p2} \\
\begin{split}
\Gamma_{015678}\epsilon_{R}&=\epsilon_{L} \, .
\end{split} \label{nsns1p1}
\end{align}
From the conditions 
(\ref{ns5p1}), (\ref{d3p1}), (\ref{nsns1p1}) and (\ref{nsns1p2}), 
there are three non-trivial projection conditions, 
so 4 supercharges are preserved in the brane configuration (\ref{nsnsbc1}). 
In order to see the chirality of the two-dimensional supersymmetry, 
we note the conditions 
\begin{align}
\label{aproj1a}
\Gamma_{26}\epsilon_{L}=-\Gamma_{59}\epsilon_{L} \, ,		\qquad		\Gamma_{26}\epsilon_{R}=-\Gamma_{59}\epsilon_{R}
\end{align}
from above brane configurations. 
Since the four-dimensional world-volume of the D3-branes is finite along $x^{6}$ 
and the effective field theory is three-dimensional, 
we may treat $\Gamma_{6}$ essentially proportional to the identity matrix. 
Here $\Gamma_{2}$ plays the role of the 
two-dimensional chirality matrix for the two-dimensional boundary 
of the three-dimensional field theory 
while $\Gamma_{5}$ (resp. $\Gamma_{9}$) is chirality matrix for the $SO(2)_{34}$ (resp. $SO(2)_{78}$). 
Let $(\pm,\pm,\pm)$ be the representation under 
the $SO(1,1)\times SO(2)_{34}\times SO(2)_{78}$ 
where $\pm$ denote the two-dimensional chiralities. 
Suppose that chiral supersymmetry is preserved at the two-dimensional boundary, 
say the right-moving $(+,\cdot,\cdot)$ supersymmetry. 
As the $SO(2)_{34}$ charge and the $SO(2)_{78}$ charge are constrained via \eqref{aproj1a},
we would only have 2 supercharges with $(+,+,-)$ and $(+,-,+)$ 
if we choose positive multiplicative constant for $\Gamma_6$, which we treated as the identity matrix.
However, since we have 4 supercharges in the brane setup (\ref{nsnsbc1}), 
this implies that there should also be left-moving supersymmetry. 
Therefore the additional NS5$'$- and D5$'$-branes preserve
the non-chiral $\mathcal{N}=(2,2)$ supersymmetry where $SO(2)_{34}\times SO(2)_{78} \cong U(1)_{\text{axial}} \times U(1)_{\text{vector}}$ are axial and vector R-symmetry of the 2d $\mathcal{N}=(2,2)$ supersymmetry.

\begin{itemize}[leftmargin=5mm]

\item[i)] NS5$'$-brane

The D3-branes ending on the NS5$'$-brane can fluctuate along $x^{3},x^{4}$ 
and the two-dimensional gauge field $A_{m}$ can fluctuate 
at the boundary. 
On the other hand, 
the NS5$'$-brane gives Dirichlet boundary condition for $A_2$ also for $\phi^5$ as it is localized at $x^5$.
These boundary conditions are consistent with 
the electric-like A-type boundary conditions 
(\ref{asusybc3c})-(\ref{asusybc3e})
\begin{align}
\label{bc1nsa1}
\begin{array}{ll}
F_{2m}=0&\textrm{(Neumann-like)}		\\
D_{2}\phi^{a}=0&\textrm{(Neumann-like)}		\\
D_{m}\phi^{5}=0&\textrm{(Dirichlet-like)} \, .\\
\end{array}
\end{align}

\item[ii)] D5$'$-brane

As $x^3$ and $x^4$ position of the D3-branes are fixed by the D5$'$-brane  
but the motion of the D3-brane along $x^5$ is unconstrained, 
$\phi^3$ and $\phi^4$ satisfy Dirichlet-like condition but $\phi^5$ would satisfy Neumann-like condition. 
The boundary condition for the two-dimensional gauge field $A_{m}$ 
imposed by the D5$'$-brane is the Dirichlet-like boundary condition. 
Therefore inserting D5$'$-brane would give 
\begin{align}
\label{bc1d5a1}
\begin{array}{ll}
F_{01}=0&\textrm{(Dirichlet-like)}			\\
D_{2}\phi^{5}=0&\textrm{(Neumann-like)}		\\
D_{m}\phi^{a}=0&\textrm{(Dirichlet-like)} \, .	
\end{array}
\end{align}
In addition, the attached D5$'$-brane can leave $A_{2}$ unconstrained. This is consistent with field theoretic analysis in section \ref{secvm1}.

\end{itemize}

\subsubsection{B-type boundary conditions}
\label{secd3nsnsb}

\begin{figure}[h]
\centering
    \begin{tabular}{cc}
          \includegraphics[scale=0.5]{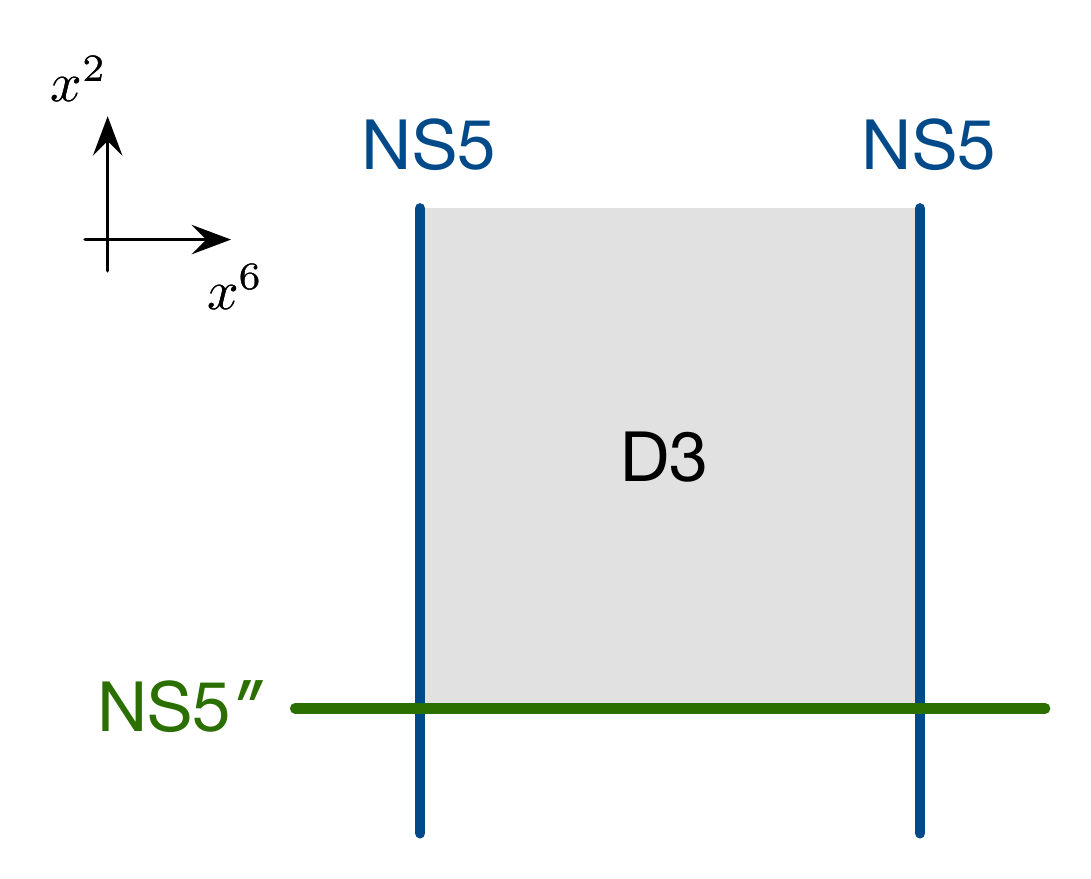}&\includegraphics[scale=0.5]{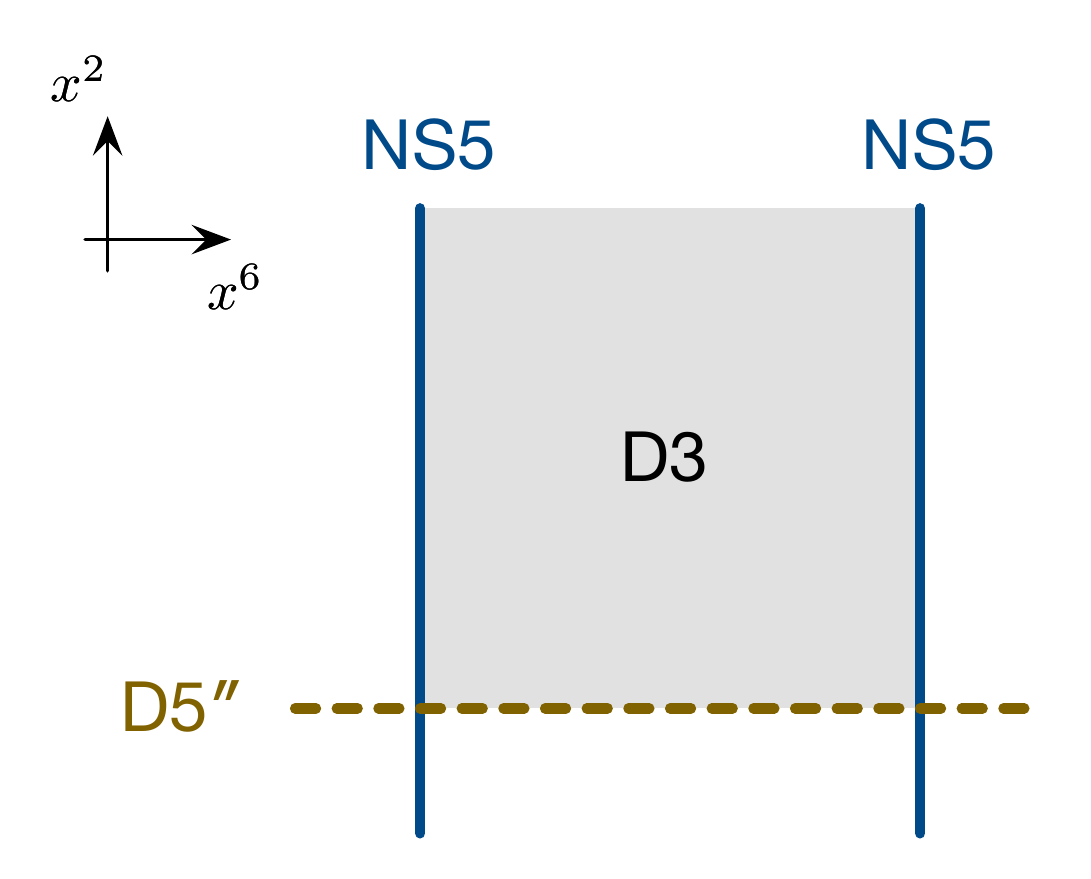}\\
          (a)NS5$''$-brane&(b)D5$''$-brane
\end{tabular}
    \caption{D3-NS5 system with NS5$''$-brane or D5$''$-brane. 
    NS5$''$-(D5$''$-)brane provides the electric-(magnetic-)like B-type boundary conditions for the vector multiplet where 2d $\mathcal{N}=(0,4)$ vector multiplet (twisted hypermultiplet) can fluctuate at the boundary.    	
    }
    \label{fig30}
\end{figure}

There are other additional 5-branes which can preserve 
$\mathcal{N}=(0,4)$ supersymmetry at the two-dimensional boundary of the 3d effective theories. 
We consider the NS5$''$-brane with world-volume $(x^{0},x^{1},x^{6},x^{7},x^{8},x^{9})$ 
or the D5$''$-brane with world-volume $(x^{0},x^{1},x^{3},x^{4}, x^5, x^{6})$ located at $x^{2}=0$ 
where D3-branes are extended on the half-space $x^2 \geq 0$ (see Figure \ref{fig30});
\begin{align}
\label{nsnsbc2}
\begin{array}{ccccccccccc}
&0&1&2&3&4&5&6&7&8&9\\
\textrm{D3}
&\circ&\circ&\circ&-&-&-&\circ&-&-&- \\
\textrm{NS5}
&\circ&\circ&\circ&\circ&\circ&\circ&-&-&-&- \\
\textrm{NS5$''$}
&\circ&\circ&-&-&-&-&\circ&\circ&\circ&\circ \\
\textrm{D5$''$}
&\circ&\circ&-&\circ&\circ&\circ&\circ&-&-&- \\
\end{array}
\end{align}

These additional NS5$''$ and D5$''$ give constraints, respectively, 
\begin{align}
\begin{split}
\Gamma_{016789}\epsilon_{L} =\epsilon_{L}, \quad &
\Gamma_{016789}\epsilon_{R} =-\epsilon_{R}
\end{split} \label{nsns2p1}
\\
\begin{split}
\Gamma_{013456}\epsilon_{R}&=\epsilon_{L}.
\end{split} \label{nsns2p2}
\end{align}
From the conditions (\ref{nsns2p1}), (\ref{nsns2p2}), 
(\ref{ns5p1}), and (\ref{d3p1}), 
we have three non-trivial projection conditions, 
so there are 4 supercharges in the brane system (\ref{nsnsbc2}). 
Also, the set of conditions lead to
\begin{align}
\label{bproj1a}
\Gamma_{01}\epsilon_{L} = \epsilon_{L} \, ,	\qquad		\Gamma_{01}\epsilon_{R} = \epsilon_{R} \, ,
\end{align}
which implies that we have chiral $\mathcal{N}=(0,4)$ supersymmetry at the two-dimensional boundary. 

The inclusion of these additional 5-branes doesn't break the symmetry 
$SO(3)_{345} \times SO(3)_{789} \cong SU(2)_{C}\times SU(2)_{H} \cong SO(4)_{R}$, which is 
the R-symmetry 
of 2d $\mathcal{N}=(0,4)$ supersymmetry. 
Under the $SO(1,1)\times SU(2)_{C}\times SU(2)_{H}$, 
the preserved right-moving supercharges transform as $(\bm{2},\bm{2})_{+}$.

\begin{itemize}[leftmargin=5mm]

\item[i)] NS5$''$-brane

The NS5$''$-brane fixes the motion of the D3-branes in $x^{3}, x^{4},x^{5}$,  
so three scalar fields $\phi^{i}$ obey the Dirichlet-like boundary conditions. 
On the other hand,
the two-dimensional gauge field $A_{m}$ can fluctuate at the boundary, and $A_2$ satisfy Dirichlet-like condition. 
Therefore the NS5$''$-brane imposes the boundary conditions
\begin{align}
\label{bc1nsb1}
\begin{array}{ll}
F_{2m}=0&\textrm{(Neumann-like)}\\
D_{m}\phi^{i}=0&\textrm{(Dirichlet-like)} \; ,\\
\end{array}
\end{align}
which are consistent with NS5$''$-like B-type boundary conditions 
(\ref{bsusybc3a}) and (\ref{bsusybc3b}).

\item[ii)] D5$''$-brane

Since D5$''$-brane is extended along $x^{3},x^{4},x^{5}$, 
the three scalar fields $\phi^{i}$ can free to move at the boundary. 
They transform as $(\bm{3,\bm{1}})$ under $SO(3)_{345}\times SO(3)_{789}$. 
Meanwhile, the two-dimensional gauge field $A_{m}$ satisfies Dirichlet condition
because it is tangent to the D5$''$-brane, 
but the scalar field $A_{2}$ can fluctuate at the boundary. 
Thus, for a single D3-brane, the D5$''$-brane would give the boundary conditions 
\begin{align}
\label{bc1d5b0}
\begin{array}{ll}
F_{01}=0&\textrm{(Dirichlet-like)}\\
D_{2}\phi^{i}=0&\textrm{(Neumann-like)}.\\
\end{array}
\end{align}
However, considering the field theory result discussed in section \ref{secvm1}, 
we expect that the above boundary condition is generalized to 
\begin{align}
\label{bc1d5b1}
\begin{array}{ll}
F_{01}=0&\textrm{(Dirichlet-like)}		\\
D_{2}\phi^{i} -\frac{1}{2} i \epsilon^{ijk}[\phi^{j},\phi^{k}]
=0&\textrm{(Nahm-like)} \, .	
\end{array}
\end{align}
That is, we expect that D3-NS5-D5$''$ realize the magnetic-like B-type boundary conditions, which are described by \eqref{bc1d5b1} including Nahm-like equation.
This is reminiscent of appearance of Nahm equation 
in half-BPS boundary conditions of 4d $\mathcal{N}=4$ theories 
discussed in \cite{Gaiotto:2008sa} 
where the nontrivial boundary conditions for multiple stack of D3-branes provided by D5-brane are described 
by Nahm equation due to the existence of the fluctuating scalar fields $A_2$.

\end{itemize}

\subsection{D3-D5 branes}
\label{secd3d5d5}
Next we consider the $N$ D3-branes suspended between the two parallel D5-branes. 
In the low-energy limit, the world-volume theory of the D3-branes is 
a theory of $N$ massless 3d $\mathcal{N}=4$ hypermultiplets \cite{Hanany:1996ie}. 
The bosonic massless modes in the theories 
are the fluctuations of the D3-branes in transverse positions $x^{7},x^{8},x^{9}$ 
which we will denote by $X^{\hat{7}},X^{\hat{8}},X^{\hat{9}}$ 
and the scalar field $A_{6}$. 
They combine into two complex scalar fields 
transforming as $(\bm{1},\bm{2})$ under $SU(2)_{C}\times SU(2)_{H}$. 
The mass parameters, $\left\{ {M}, \phi_{M} \right\}$, 
are given by the relative position of the D5-branes along $\{x^3,x^4, x^5\}$.

\subsubsection{A-type boundary conditions}
\label{secd3d5d5a}

\begin{figure}[h]
\centering
    \begin{tabular}{cc}
          \includegraphics[scale=0.5]{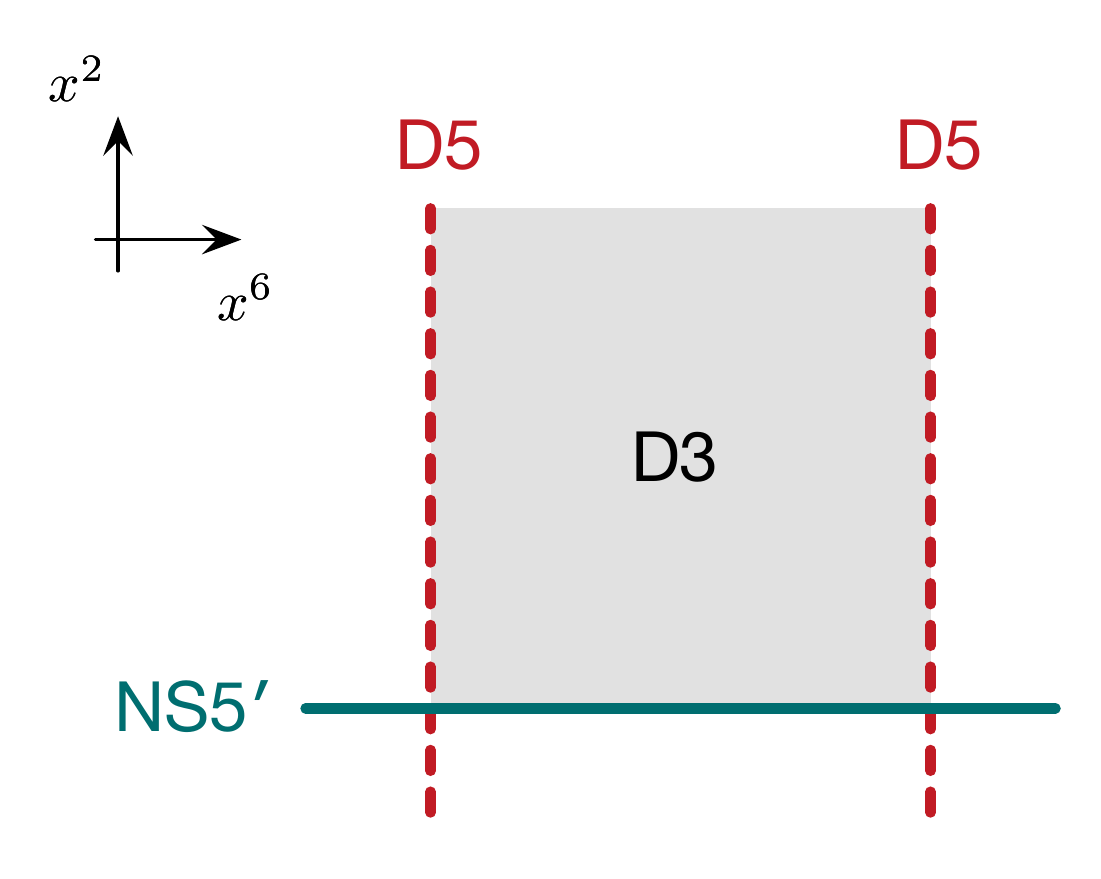}&\includegraphics[scale=0.5]{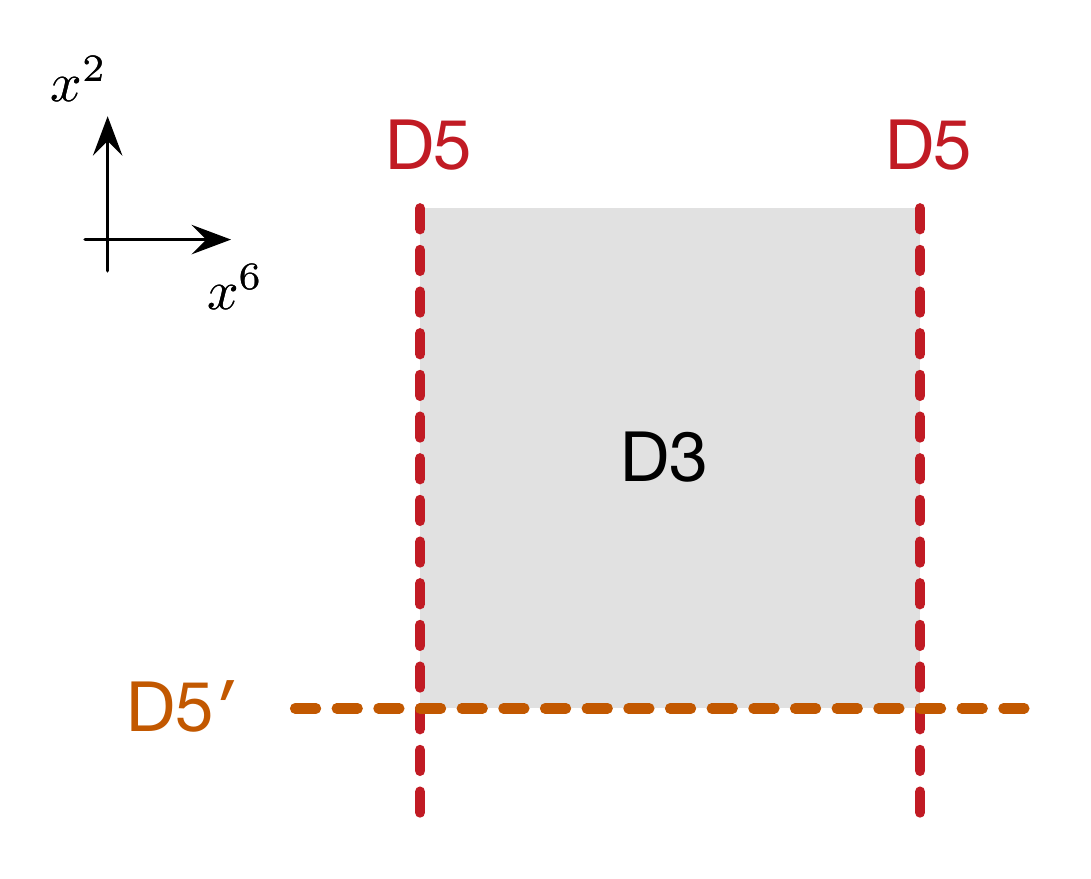}\\
          (a)NS5$'$-brane&(b)D5$'$-brane
\end{tabular}
    \caption{D3-D5 system with NS5$'$-brane or D5$'$-brane. 
    NS5$'$ and D5$'$-brane provide the A-type boundary conditions for the pure hypermultiplet where 2d $\mathcal{N}=(2,2)$ chiral multiplets can fluctuate at the boundary.
    }
    \label{fig20}
\end{figure}

As discussed in subsection \ref{secd3nsns}, 
we can realize the two-dimensional non-chiral $\mathcal{N}=(2,2)$ supersymmetry 
by the introduction of the NS5$'$- or D5$'$-branes 
\begin{align}
\label{d5d5nsa}
\begin{array}{ccccccccccc}
&0&1&2&3&4&5&6&7&8&9\\
\textrm{D3}
&\circ&\circ&\circ&-&-&-&\circ&-&-&- \\
\textrm{D5}
&\circ&\circ&\circ&-&-&-&-&\circ&\circ&\circ \\
\textrm{NS5$'$}
&\circ&\circ&-&\circ&\circ&-&\circ&-&-&\circ \\
\textrm{D5$'$}
&\circ&\circ&-&-&-&\circ&\circ&\circ&\circ&-\\
\end{array}
\end{align}
as in the configuration (\ref{nsnsbc1}) (see Figure \ref{fig20}).

Under the space-time symmetry $SO(1,1)\times SO(2)_{34}\times SO(2)_{78}$ 
$\cong$ $SO(1,1)\times U(1)_{C}\times U(1)_{H}$, 
the three scalar fields $X^{\hat{i}}$, $\hat{i}=7,8,9$ are divided into 
the two scalar fields $X^{\hat{a}}$, $\hat{a}=7,8$ 
and the scalar field $X^{\hat{9}}$. 
As $SU(2)_{H}$ is broken to $U(1)_{H}$, 
these scalar fields are charged under the vector R-symmetry 
of 2d $\mathcal{N}=(2,2)$ theories.

\begin{itemize}[leftmargin=5mm]

\item[i)] NS5$'$-brane

As the D3-branes can move along $x^9$ in the presence of NS5$'$, 
the scalar field $X^{\hat{9}}$, which describes the position of the D3-branes along $x^9$, can fluctuate at the boundary.
In addition, 
the massless modes of the scalar field $A_{6}$ 
can also fluctuate as the NS5$'$-brane is extended along $x^{6}$. 
Thus the additional NS5$'$-brane keeps the half of the bosonic degrees of freedom of 
the 3d $\mathcal{N}=4$ hypermultiplet at the boundary, 
\begin{align}
\label{bc2nsa0}
\begin{array}{lll}
\partial_{m}X^{\hat{7}}=0,
&\partial_{m}X^{\hat{8}}=0&\textrm{(Dirichlet-like)}\\
\partial_{2}X^{\hat{9}}=0,
&\partial_{2}A_{6}=0&\textrm{(Neumann-like)}	\, .\\
\end{array}
\end{align}
Let $q = X^{\hat{7}}+iA_{6}$ and $\widetilde{q}=X^{\hat{8}}+iX^{\hat{9}}$ be two complex scalar fields. 
Then we have
\begin{align}
\label{bc2nsa01}
\begin{array}{lll}
\partial_{m}(\operatorname{Re}q)=0,			&\partial_{m}(\operatorname{Re}\widetilde{q})=0		&\textrm{(Dirichlet-like)}\\
\partial_{2}(\operatorname{Im}q)=0,			&\partial_{2}(\operatorname{Im}\widetilde{q})=0		&\textrm{(Neumann-like)}	\, .\\
\end{array}
\end{align}

\item[ii)] D5$'$-brane

As D3-branes can move along $x^7$ and $x^8$ directions, scalar fields $X^{\hat{7}}$ and $X^{\hat{8}}$ corresponding to directions $x^7$ and $x^8$ can fluctuate at the boundary.  
On the other hand, the scalar field $X^{\hat{9}}$ corresponding to $x^9$ cannot fluctuate at the boundary.
Also, the massless modes associated to $A_{6}$ 
cannot fluctuate at the boundary since D5$'$ is extended along $x^6$.
Similarly to the case with the NS5$'$-brane, the half of the bosonic degrees of freedom of 
the 3d $\mathcal{N}=4$ hypermultiplet can survive at the boundary. 
Therefore we have
\begin{align}
\label{bc2d5a0}
\begin{array}{lll}
\partial_{2}X^{\hat{7}}=0,
&\partial_{2}X^{\hat{8}}=0&\textrm{(Neumann-like)}\\
\partial_{m}X^{\hat{9}}=0,
&\partial_{m}A_{6}=0&\textrm{(Dirichlet-like)} \, .\\
\end{array}
\end{align}
Again, in terms of $q$ and $\widetilde{q}$ we have 
\begin{align}
\label{bc2d5a01}
\begin{array}{lll}
\partial_{2}(\operatorname{Re}q)=0,		&\partial_{2}(\operatorname{Re}\widetilde{q})=0		&\textrm{(Neumann-like)}\\
\partial_{m}(\operatorname{Im}q)=0,	&\partial_{m}(\operatorname{Im}\tilde{q})=0	&\textrm{(Dirichlet-like)} \, .\\
\end{array}
\end{align}

\end{itemize}

\subsubsection{B-type boundary conditions}
\label{secbbrane1a}
\begin{figure}[h]
\centering
    \begin{tabular}{cc}
          \includegraphics[scale=0.5]{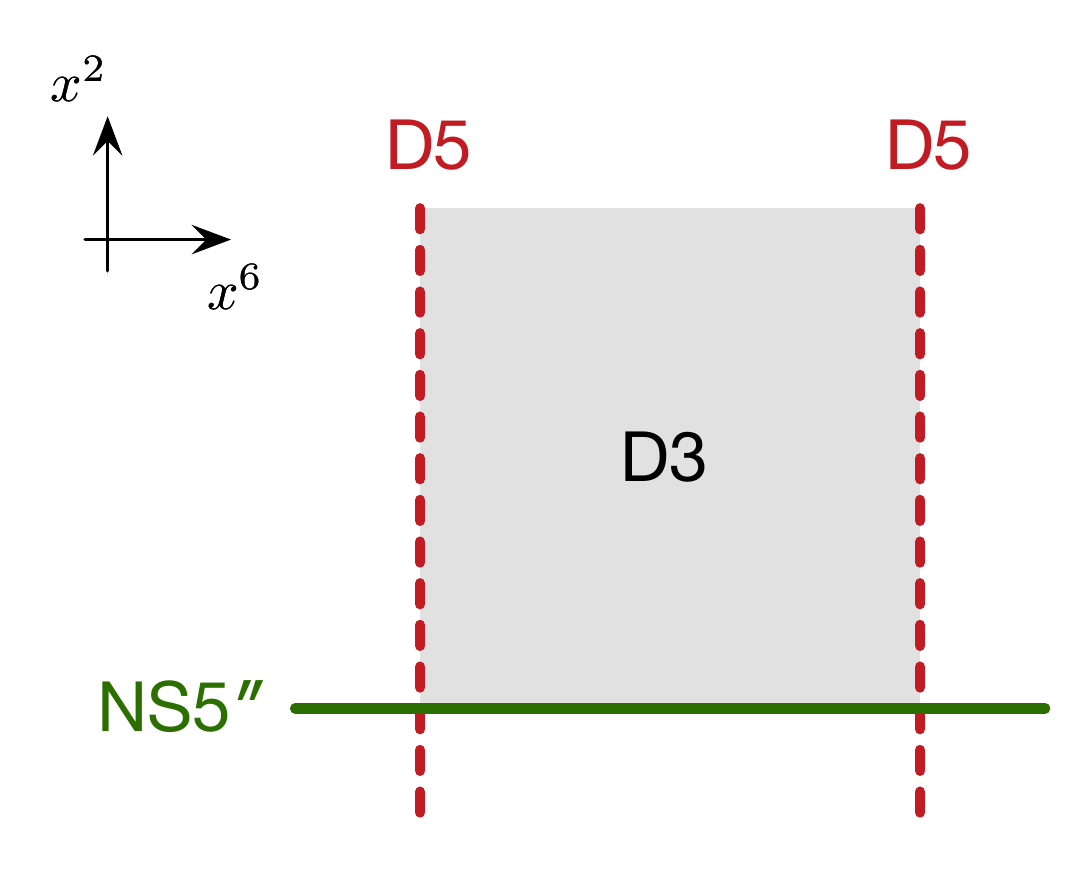}&\includegraphics[scale=0.5]{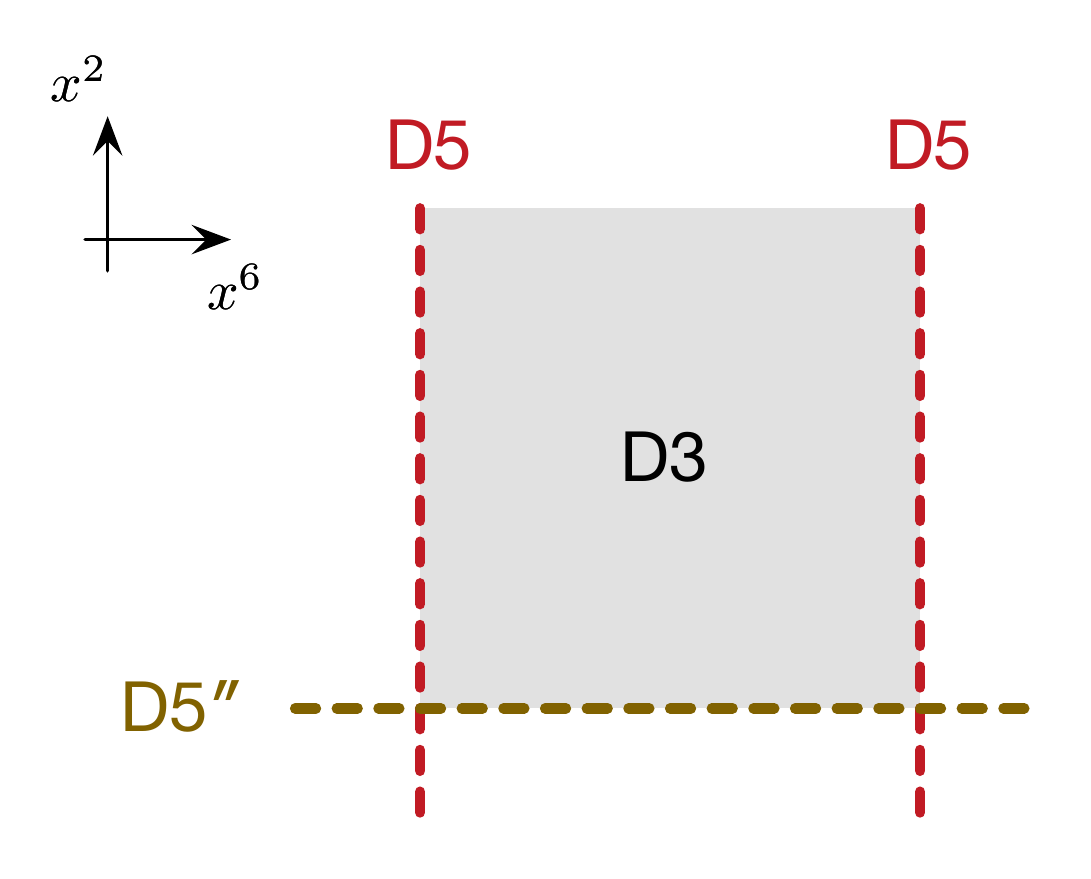}\\
          (a)NS5$''$-brane&(b)D5$''$-brane
\end{tabular}
    \caption{D3-D5 system with NS5$''$-brane or D5$''$-brane. 
    NS5$''$-(D5$''$-)brane provides the B-type boundary conditions for the pure hypermultiplet where 2d $\mathcal{N}=(0,4)$ hypermultiplet ($\mathcal{N}=(0,4)$ Fermi multiplet) can fluctuate at the boundary.
    }
    \label{fig40}
\end{figure}

Following the arguments for the D3-NS5 brane system, 
$\mathcal{N}=(0,4)$ supersymmetry can be preserved at the boundary
by adding the NS5$''$- or D5$''$-branes at $x^2=0$ to the D3-D5 brane configuration 
where D3-branes are extended along $x^2 \geq 0$ (see Figure \ref{fig40}) as
\begin{align}
\label{d5d5d5b}
\begin{array}{ccccccccccc}
&0&1&2&3&4&5&6&7&8&9\\
\textrm{D3}
&\circ&\circ&\circ&-&-&-&\circ&-&-&- \\
\textrm{D5}
&\circ&\circ&\circ&-&-&-&-&\circ&\circ&\circ \\
\textrm{NS5$''$}
&\circ&\circ&-&-&-&-&\circ&\circ&\circ&\circ \\
\textrm{D5$''$}
&\circ&\circ&-&\circ&\circ&\circ&\circ&-&-&- \\
\end{array}
\end{align}

The brane configuration (\ref{d5d5d5b}) preserves  
R-symmetry $SO(4)_R = SU(2)_C \times SU(2)_H \cong SO(3)_{345} \times SO(3)_{789}$ of 3d $\mathcal{N}=4$ theories and 
the three scalar fields $X^{\hat{i}}$ transform 
as a triplet under $SO(3)_{789}$.

\begin{itemize}[leftmargin=5mm]

\item[i)] NS5$''$-brane

Since NS5$''$-brane is supported on $x^6, x^{7}, x^{8}$, and $x^{9}$ directions, 
scalar field $A_{6}$ and three scalar fields $X^{\hat{i}}$ can fluctuate at the boundary. 
The NS5$''$-brane would lead to the Neumann conditions for these scalar fields
\begin{align}
\label{bc2nsb0}
\partial_{2}X^{\hat{i}} = 0 \, ,	\qquad	\partial_{2}A_{6} = 0 \, .
\end{align}
These conditions correspond to Neumann boundary conditions (\ref{hbsusybc3b2})
for the pure hypermultiplets 
\begin{align}
\label{bc2nsb01}
\begin{array}{lll}
\partial_{2}q = 0 \, , 	\qquad	 \partial_{2}\widetilde{q}=0 \qquad \textrm{(Neumann-like)} 
\end{array}
\end{align}

\item[ii)] D5$''$-brane

Since the D5$''$-brane is extended in $x^{6}$ and located at $x^{7}, x^{8}, x^{9}$, 
the scalar field $A_{6}$ and the three scalar fields describing the position of the D3-branes 
all satisfy Dirichlet condition at the boundary,
\begin{align}
\label{bc2d5b0}
\partial_{m}X^{\hat{i}} = 0 \, ,  	\qquad 	\partial_{m}A_{6} = 0 \, .
\end{align}
We see that 
the above conditions (\ref{bc2d5b0}) are equivalent to the conditions (\ref{hbsusybc3a2}).

Hence, in terms of $q$ and $\widetilde{q}$, the conditions read
\begin{align}
\label{bc2d5b1}
\begin{array}{lll}
\partial_{m}q = 0 \, , 	\qquad 		\partial_{m}\widetilde{q} = 0 \qquad 	\textrm{(Dirichlet-like)}.
\end{array}
\end{align}

\end{itemize}

\subsection{D3-NS5-D5 branes}
\label{secd3nsd5}
We consider A- and B-type boundary conditions for SQCD in the context of brane configuration \eqref{d3nsd5}.

\subsubsection{A-type boundary conditions}
\label{secd3nsd5a}
Similarly as before, we consider the extra NS5$'$- or D5$'$-branes at $x^2=0$ in the following brane configurations,
\begin{align}
\label{d3ns5d5A}
\begin{array}{ccccccccccc}
&0&1&2&3&4&5&6&7&8&9\\
\textrm{D3}
&\circ&\circ&\circ&-&-&-&\circ&-&-&- \\
\textrm{NS5}
&\circ&\circ&\circ&\circ&\circ&\circ&-&-&-&- \\
\textrm{D5}
&\circ&\circ&\circ&-&-&-&-&\circ&\circ&\circ \\
\textrm{NS5$'$}
&\circ&\circ&-&\circ&\circ&-&\circ&-&-&\circ \\
\textrm{D5$'$}
&\circ&\circ&-&-&-&\circ&\circ&\circ&\circ&-\\
\end{array}
\end{align}
As usual, 3d $\mathcal{N}=4$ vector multiplet is realized in the world-volume of D3-branes. 
Also the hypermultiplet is realized as strings connecting D3-branes and D5-branes. 
We expect that when NS5$'$-(D5$'$-)brane is added, 
it provides Neumann-(Dirichlet-)like condition 
for $\{A_{m}, \phi^5 \}$ and $\{ \text{Im}(q), \text{Im}(\widetilde{q}) \}$, 
but Dirichlet-(Neumann-)like condition 
for $\{ A_2, \phi^a \}$ and $\{ \text{Re}(q), \text{Re}(\widetilde{q}) \}$ where ${m}=0,1$, $a=3,4$.

Since NS5$'$-brane is located at $x^{7}$, $x^{8}$, 
two of the FI parameters of 3d $\mathcal{N}=4$ theory would arise in the boundary conditions 
as the relative positions of NS5 branes in $x^{7}$, $x^{8}$ directions. 
This brane picture is consistent to the result that 
the deformed A-type boundary conditions (\ref{dhasusybc1d}) for the coupled hypermultiplets
involve the two FI parameters $r$ and $\operatorname{Re} \phi_{r}$. 
In this brane configuration, mass parameters of 3d $\mathcal{N}=4$ theory are given by
the relative distance between D3 and D5 branes in $x^{3}$, $x^{4}$, $x^{5}$ directions. 
The mass parameter $\operatorname{Im} \phi_{M}$, which generalizes 
the hypermultiplet Neumann boundary conditions (\ref{hasusybc2b}) 
to the Robin-type boundary conditions (\ref{dhasusybc2c}) is given by 
the relative distance between D3-branes and D5-branes along $x^5$ direction where the position of D3-branes along $x^5$ direction is fixed by NS5$'$-brane.
Meanwhile, the mass parameter ${M}$, which is related to the vev of $\phi^3$ of background vector multiplet, has different nature from $\text{Im}\phi_{M}$ above. 
Given the position of D5 brane at fixed location of $x^3$ and $x^4$ directions, since NS5$'$ brane is supported on $x^3$ and $x^4$ directions, D3 brane can still move along those directions. This is compatible with the BPS equations (\ref{dhasusybc2c0}) that mass parameter $M$ appears in boundary coupling rather than boundary conditions.

For the D5$'$-brane, since it is located at $x^9$ direction,
the boundary conditions would be deformed by one of the FI parameters of 3d $\mathcal{N}=4$ theory 
as the relative position of NS5 branes in $x^9$ direction.  
In field theory analysis, we see that a single FI parameter $\operatorname{Im} \phi_{r}$ appears 
in the deformed A-type boundary conditions (\ref{dhasusybc2d}) for the coupled hypermultiplets. 
As D5$'$-brane is located at $x^{3}$, $x^{4}$ directions, in a similar manner discussed above, 
two mass parameters would generalize the hypermultiplet boundary conditions. 
Those corresponding two mass parameters $M$ and $\operatorname{Im} \phi_{M}$ 
appear in the deformed hypermultiplet boundary conditions (\ref{dhasusybc1a}) and (\ref{dhasusybc1c0}).

\subsubsection{B-type boundary conditions}
\label{secd3nsd5b}

Also, we consider the extra NS5$''$- or D5$''$-branes at $x^2=0$ to the following brane configurations,
\begin{align}
\label{d3ns5d5B}
\begin{array}{ccccccccccc}
&0&1&2&3&4&5&6&7&8&9\\
\textrm{D3}
&\circ&\circ&\circ&-&-&-&\circ&-&-&- \\
\textrm{NS5}
&\circ&\circ&\circ&\circ&\circ&\circ&-&-&-&- \\
\textrm{D5}
&\circ&\circ&\circ&-&-&-&-&\circ&\circ&\circ \\
\textrm{NS5$''$}
&\circ&\circ&-&-&-&-&\circ&\circ&\circ&\circ \\
\textrm{D5$''$}
&\circ&\circ&-&\circ&\circ&\circ&\circ&-&-&- \\
\end{array}
\end{align}

Similarly as before, we expect that when NS5$''$-(D5$''$-)brane is added, 
it provides Neumann-(Dirichlet-)like condition for $\{A_{m} \}$ and $\{q, \widetilde{q} \}$, 
but Dirichlet-(Neumann-)like condition for $\{ A_6, \phi^i \}$.

As NS5$''$-brane is supported on $x^{7}$, $x^{8}$, $x^{9}$ directions, 
none of the FI parameters of 3d $\mathcal{N}=4$ theory would deform the boundary conditions. 
This brane perspective is consistent with the result that 
the deformed B-type boundary conditions (\ref{dhbsusybc3b2}) for the coupled hypermultiplets involve no FI parameters in the boundary conditions. 
As NS5$''$-brane is located at $x^{3}$, $x^{4}$, and $x^{5}$ directions, in a similar manner discussed above, 
three mass parameters (${M}, \phi_{M}$) would appear in boundary conditions. 
In field theory analysis, we see that the hypermultiplet Neumann boundary conditions (\ref{hbsusybc3b2}) 
is generalized to the Robin-type boundary conditions (\ref{dhbsusybc3b2}) 
by all the three mass parameters ${M}^{i}$.

In the case of the D5$''$-brane, which is located at $x^{7}$, $x^{8}$, $x^{9}$ directions,
all the three FI parameters $r^{\hat{i}}$ would deform the boundary conditions. 
This can be seen from 
the deformed B-type boundary conditions (\ref{dhbsusybc3a3}) for the coupled hypermultiplets. 
Since the D5$''$-brane is supported in $x^{3}$, $x^{4}$, $x^{5}$ directions, 
mass parameters would not appear in boundary conditions. 
In fact, the deformed B-type hypermultiplet boundary conditions (\ref{dhbsusybc3a2}) and (\ref{dhbsusybc3a3}) 
are not affected by mass parameters.

\subsection{$S$-duality}
\label{secsdual}

From the analysis on the half-BPS boundary condition of 3d $\mathcal{N}=4$ theory, 
we saw that which 2d supermultiplet of $\mathcal{N}=(2,2)$ and $\mathcal{N}=(0,4)$ from bulk 3d $\mathcal{N}=4$ multiplet arise at the boundary. 
We also found that such boundary condition can be consistently understood in terms of brane configurations of Type IIB string theory. 

Upon $S$-duality of Type IIB string theory, the 3d $\mathcal{N}=4$ theory 
arising from a given brane configuration enjoy the mirror symmetry \cite{Hanany:1996ie}. 
With additional branes that provide the half-BPS boundary condition discussed in previous sections, it is interesting to see the relation between the boundary degrees of freedom arising from a particular brane configuration and those arising from $S$-dual configuration of the original brane configuration. 
In general, this could be nontrivial task, but here we just take the simplest cases, pure vector multiplet and pure hypermultiplet, discussed in previous section, and would like to see how the boundary degree of freedom from the bulk 3d $\mathcal{N}=4$ multiplet are mapped to each other.

\subsubsection{A-type}

In this case, we have $U(1)_C \times U(1)_H$ R-symmetry of 2d
$\mathcal{N}=(2,2)$ theory, which is the axial and vector R-symmetry,
from original $SU(2)_C \times SU(2)_H$ R-symmetry of 3d
$\mathcal{N}=4$. As $SU(2)_C$ and $SU(2)_H$ are exchanged under $RS$
map\footnote{
In brane configuration of Type IIB string theory, 
$R$ of $RS$ map denotes the the map
$x^{i}$ to $x^{i+4}$ where $i=3,4,5$ and $S$ denotes $S$-duality 
\cite{Hanany:1996ie}. 
In the followings, we mean $S$-duality by $RS$-duality.}, 
so $U(1)_C$ and $U(1)_H$ are exchanged. 
Hence it is expected that 
3d $\mathcal{N}=4$ mirror symmetry is closely related to 
2d $\mathcal{N}=(2,2)$ mirror symmetry through $S$-duality in Type IIB string theory. 
In fact, it has been argued that 
3d mirror symmetry decends to 2d mirror symmetry 
via compactification \cite{Aganagic:2001uw} and also that 2d $\mathcal{N}=(2,2)$ interface theory between 3d $\mathcal{N}=4$ mirror pairs 
produce mirror map of 2d $\mathcal{N}=(2,2)$ 
chiral and twisted chiral operators \cite{Bullimore:2016nji}. 
We see for
the following simplest example 
that the 2d mirror map is realized as $S$-duality in Type IIB string theory.

\begin{itemize}[leftmargin=5mm]

\item[i)] D3-NS5-NS5$'$ $\stackrel{S\text{-dual}}{\longleftrightarrow}$ D3-D5-D5$'$

The boundary degree of freedom from the bulk 3d $\mathcal{N}=4$ vector multiplet arising in D3-NS5-NS5$'$ system is 
2d $\mathcal{N}=(2,2)$ vector multiplet or field strength multiplet, which is a twisted chiral multiplet. 
On the other hand, the one from the bulk 3d $\mathcal{N}=4$ hypermultiplet arising in D3-D5-D5$'$ system is 
2d $\mathcal{N}=(2,2)$ chiral multiplet in the adjoint representation. 
As two brane configurations are $S$-dual, which gives rise to mirror pair between the pure vector multiplet and the pure hypermultiplet in the bulk, 
we see that the twisted chiral multiplet and chiral multiplet at the boundary $x^2=0$ are exchanged under $S$-duality of Type IIB string theory 
or 3d $\mathcal{N}=4$ mirror symmetry. 
This is consistent with 2d $\mathcal{N}=(2,2)$ mirror symmetry.

\item[ii)] D3-NS5-D5$'$ $\stackrel{S\text{-dual}}{\longleftrightarrow}$ D3-D5-NS5$'$

Similarly, in this case, the boundary degree of freedom from the bulk vector multiplet arising in D3-NS5-D5$'$ system is 
2d $\mathcal{N}=(2,2)$ twisted chiral multiplet, and the one from the bulk hypermultiplet arising in D3-D5-NS5$'$ is 2d $\mathcal{N}=(2,2)$ chiral multiplet. Under $S$-duality of the brane configuration, those two 2d $\mathcal{N}=(2,2)$ supermultiplet are mapped each other, which is consistent with 2d $\mathcal{N}=(2,2)$ mirror symmetry.

\end{itemize}

\subsubsection{B-type}

The 2d $\mathcal{N}=(0,4)$ mirror symmetry has not been studied much in literature.\footnote{The 2d $\mathcal{N}=(0,4)$ mirror symmetry could be understood 
as the special case of 2d $\mathcal{N}=(0,2)$ mirror symmetry \cite{Adams:2003zy}.} 
We expect that 
the $\mathcal{N}=(0,4)$ theory arising from (more general or complicated version of) 
our brane configuration and the theory arising from the corresponding $S$-dual configuration 
give rise to the $\mathcal{N}=(0,4)$ mirror pair. 
In 2d $\mathcal{N}=(0,4)$ gauge theory may receive the anomaly 
from massless charged chiral fermions running in one-loop \cite{AlvarezGaume:1983ig} 
and we should take into account the cancelation of gauge anomaly to obtain the effective theories. 
We hope to revisit this issue in the context of brane configuration. 
Here, we only consider 
the map between the 2d $\mathcal{N}=(0,4)$ supermultiplets at the boundary arising from the 3d $\mathcal{N}=4$ pure vector multiplet and the pure  hypermultiplet discussed in previous section.

\begin{itemize}[leftmargin=5mm]

\item[i)] D3-NS5-NS5$''$ $\stackrel{S\text{-dual}}{\longleftrightarrow}$ D3-D5-D5$''$

The boundary degree of freedom from 3d $\mathcal{N}=4$ vector multiplet arising in D3-NS5-NS5$''$ system is 2d $\mathcal{N}=(0,4)$ vector multiplet, and the one from 3d $\mathcal{N}=4$ hypermultiplet arising in D3-D5-D5$''$ system is 2d $\mathcal{N}=(0,4)$ Fermi multiplet. 

The 2d $\mathcal{N}=(0,4)$ vector multiplet is made of $\mathcal{N}=(0,2)$ vector multiplet 
and Fermi multiplet in adjoint representation where the $\mathcal{N}=(0,2)$ vector multiplet 
can be expressed as $\mathcal{N}=(0,2)$ field strength multiplet, which is $\mathcal{N}=(0,2)$ Fermi multiplet. 
The fermions in $\mathcal{N}=(0,4)$ vector multiplet is charged 
under $SO(1,1) \times SU(2)_C \times SU(2)_H$ as $(\mathbf{2},\mathbf{2})_-$. 
Meanwhile, the $\mathcal{N}=(0,4)$ Fermi multiplet is made of 
two $\mathcal{N}=(0,2)$ Fermi multiplet in conjugate representation of gauge group $G$, 
and it is charged under $SU(2)_C \times SU(2)_H$ as $(\mathbf{1},\mathbf{1})_-$. 
Since there are four real fermions in the vector multiplet, under the $S$-duality of IIB theory, 
the number of fermions is matched with the number of them in Fermi multiplet 
though it is not quite sure to explain the relation of their R-charges in the scope of this paper.
It seems that better understanding is needed for this case.

\item[ii)] D3-NS5-D5$''$ $\stackrel{S\text{-dual}}{\longleftrightarrow}$ D3-D5-NS5$''$

The boundary degree of freedom from the bulk 3d $\mathcal{N}=4$ vector multiplet arising in D3-NS5-D5$''$ system is 
2d $\mathcal{N}=(0,4)$ twisted hypermultiplet, 
and the one from the 3d $\mathcal{N}=4$ hypermultiplet arising in D3-D5-NS5$''$ system 
is 2d $\mathcal{N}=(0,4)$ hypermultiplet. 
Upon $S$-duality, $SU(2)_C$ and $SU(2)_H$ are exchanged, so twisted hypermultiplet are mapped hypermultiplet, vice versa.

\end{itemize}

\section{Conclusion and discussion}
\label{seccon}
In this paper, we studied the half-BPS boundary conditions in 3d $\mathcal{N}=4$ gauge theories 
preserving $\mathcal{N}=(2,2)$ and $(0,4)$ supersymmetries at the boundary, which we call A-type and B-type, respectively. 
We calculated the BPS boundary equations for vector multiplet 
and hypermultiplet involving gauge coupling, FI and mass deformations. 
We also saw that 3d bulk supermultiplets are decomposed to the boundary 
supermultiplet of preserved supersymmetry. 
We found that the boundary BPS equations for the vector multiplet, 
in particular, 
give rise to Nahm-like equation in the magnetic-like B-type boundary conditions.
For the hypermultiplet we saw that the Neumann-like boundary conditions for scalar components of hypermultiplet 
are generalized to Robin-type boundary condition upon turning on gauge coupling and mass deformation.
We proposed brane configurations in Type IIB string theory 
realizing such $\mathcal{N}=(2,2)$ and $(0,4)$ BPS boundary conditions in 3d $\mathcal{N}=4$ theories, 
and checked that they are consistent with the analysis in field theory. 
We also saw how the boundary supermultiplets from the bulk supermultiplets are mapped under $S$-duality of Type IIB theory. 
\\

In order to study the supersymmetric vacua of 3d $\mathcal{N}=4$ gauge theory on half-space, 
it is necessary to study the BPS boundary conditions in detail. 
A notable consequence is that we get Nahm-like equation in vector multiplet boundary conditions of B-type. 
It is interesting to analyze these BPS equations in a similar way discussed in \cite{Gaiotto:2008sa} for 4d $\mathcal{N}=4$ SYM theories.

Brane realization of 2d gauge theories with $(2,2)$ and $(0,4)$ supersymmetries is one of interesting subjects.\footnote{Other brane realizations for 2d $\mathcal{N}=(2,2)$ and $(0,4)$ have been discussed in \cite{Hanany:1997vm, Tong:2014yna}, respectively.} 
In particular, there is an anomaly issue in 2d $\mathcal{N}=(0,4)$ theories and 
it would be interesting to know how such anomaly condition can arise in Type IIB string theory. 
Also, as we briefly discussed for the boundary degrees of freedom from the bulk supermultiplets, realization of 2d $\mathcal{N}=(0,4)$ theories in the brane configuration will tell us, via $S$-duality of Type IIB theory, a \textit{mirror} dual theory of a given 2d $\mathcal{N}=(0,4)$ theory from the corresponding brane configurations. 
With anomaly issue taken into account, study of mirror symmetry of 2d $\mathcal{N}=(0,4)$ theory via Type IIB $S$-duality would be one of intriguing directions.

\subsection*{Acknowledgments}
We would like to thank Yutaka Yoshida for collaboration at the early stage of this work. 
We also thank Tudor Dimofte, Kazuo Hosomichi, Rak-Kyeong Seong, Satoshi Yamaguchi, and Piljin Yi for interesting discussion. 
H.J.C appreciate the IBS Center for Geometry and Physics, the Department of Mathematics at Postech, the Institut Henri Poincar\'e, and the Simons Center for Geometry and Physics (Simons Summer Workshop 2016) for hospitality where part of this work was performed. 
T.O. is grateful to NCTS Summer Workshop on Strings and Quantum Field Theory for hospitality during the completion of this work. 
Research of T.O. is supported by MOST under the Grant No.105-2811-066. 

%
%
%
%
\appendix


\section{3d $\mathcal{N}=2$ superspace and superfields}
\label{appss}

\subsection{Spinors and superspace}
\label{appspinor}
We use the metric $\eta_{\mu\nu}=\eta^{\mu\nu}=\mathrm{diag}(-1,1,1)$
and $2\times 2$ $\gamma^{\mu}$ matrices satisfy
\begin{align}
\label{3dspace1}
\{\gamma^{\mu},\gamma^{\nu}\}=2\eta^{\mu\nu}.
\end{align}
$\gamma^{0}$ is taken as anti-hermitian and $\gamma^{1}$ and
$\gamma^{2}$ as hermitian. 
We introduce a three-dimensional charge conjugation matrix $\epsilon$, 
which have the following properties: 
\begin{align}
\label{3dg1c1}
\epsilon^{\dag}&=\epsilon^{-1},&
\epsilon^{T}&=-\epsilon,&
(\epsilon\gamma^{\mu})^{T}&=\epsilon\gamma^{\mu}.
\end{align}
Two-component spinors $\psi^{\alpha}$ with upper or lower indices transform as
\begin{align}
\label{3dspinor1a}
\psi_{\alpha}&:=\epsilon_{\alpha\beta}\psi^{\beta},&
 \psi^{\alpha}&=(\epsilon^{-1})^{\alpha\beta}\psi_{\beta}.
\end{align}
We use the following summation convention:
\begin{align}
\label{3dspinor1b}
(\chi\psi)&:=\chi^{\alpha}\psi_{\alpha}
=\chi^{\alpha}\epsilon_{\alpha\beta}\psi^{\beta},&
(\gamma^{\mu}\psi)^{\alpha}
&=\gamma^{\mu}{}^{\alpha}{}_{\beta}\psi^{\beta},&
(\epsilon\gamma^{\mu}\psi)_{\alpha}
&=(\epsilon\gamma^{\mu})_{\alpha\beta}\psi^{\beta}.
\end{align} 
We define $\sigma$-matrices as 
\begin{align}
\label{3dspinor1c}
\sigma^{\mu}&:=\epsilon\gamma^{\mu},
\end{align}
and use the summation expression
$\xi\sigma^{\mu}\psi:=\xi^{\alpha}(\epsilon\gamma^{\mu})_{\alpha\beta}\psi^{\beta}$. 
We define the conjugation by
\begin{align}
\label{3dspinor1d}
\psib_{\alpha} & := - \psi^{\dagger}_{\beta} (\gamma^0)^{\beta}_{\; \alpha} \, .
\end{align}

Here are useful spinor formulae:
\begin{align}
\label{3dspinor2a}
\xi \psi&=\psi\xi,& 
\xi \sigma^{\mu} \psi&=-\psi \sigma^{\mu} \xi ,\nonumber  \\
\psi \sigma^{\mu}\psi&=0,&
\psi \epsilon\gamma^{\mu\nu}\chi&=- \chi \epsilon\gamma^{\mu\nu}\psi,
\\
\label{3dspinor2b}
(\xi \psi)^{\dag}&=-\psib\xib,&
(\xi \sigma^{\mu}\psi)^{\dag}&=\psib \sigma^{\mu}\xib=-\xib \sigma^{\mu}\psib,
\\
\label{3dspinor2c}
\theta_{\alpha}\theta_{\beta}&=\frac12 \epsilon_{\alpha \beta}\theta\theta,&
\theta^{\alpha}\theta^{\beta}&=-\frac12 (\epsilon^{-1})^{\alpha\beta}\theta\theta,
\end{align}
\begin{align}
\label{3dspinor2d}
(\theta\psi)(\theta\chi)&=-\frac12(\theta\theta)(\psi\chi) ,\\
\label{3dspinor2e}
(\theta\sigma^{\mu}\chi)(\theta \psi)
&=-\frac12\theta\theta\psi\sigma^{\mu}\chi ,\\
\label{3dspinor2f}
\theta\sigma^{\mu}\thetab \theta \sigma^{\nu}\thetab&=\frac12
\theta\theta\thetab\thetab \eta^{\mu\nu},
\end{align}
\begin{align}
\label{3dspinor2g}
-\frac12(\chi\lambda)(\psi\xi)
-\frac12(\chi\sigma^{\mu}\lambda)(\psi\sigma_{\mu}\xi)&=
(\chi\xi)(\psi\lambda),
\end{align}
where $\psi, \xi, \theta, \lambda$ are two-component spinors.

We consider the 3d $\Ncal=2$ superspace coordinates 
$(x^{\mu},\theta^{\alpha},\thetab^{\alpha})$ which transform as
$x^{\mu}\rightarrow x^{\mu}-i\epsilon\sigma^{\mu}\thetab
 -i\bar{\epsilon}\sigma^{\mu}\theta$, $\theta\rightarrow \theta+\epsilon$
and $\thetab\rightarrow \thetab+\bar{\epsilon}$
 under the supersymmetry transformations. 
Let us define the following supersymmetric derivatives:
\begin{align}
\label{n2sspace1a}
 Q_{\alpha}&:=\deldel{}{\theta^{\alpha}}-i(\sigma^{\mu}\thetab)_{\alpha}\del_{\mu},&
 \Qb_{m}&:=-\deldel{}{\thetab^{\alpha}}+i(\sigma^{\mu}\theta)_{\alpha}\del_{\mu},\\
 \label{n2sspace1b}
 D_{\alpha}&:=\deldel{}{\theta^{\alpha}}+i(\sigma^{\mu}\thetab)_{\alpha}\del_{\mu},&
 \Db_{\alpha}&:=-\deldel{}{\thetab^{\alpha}}-i(\sigma^{\mu}\theta)_{\alpha}\del_{\mu}.
\end{align}
They have the anti-commutation relations
\begin{align}
\label{n2sspace2}
 \{Q_{\alpha},\Qb_{\beta}\}&=2i\sigma^{\mu}_{\alpha\beta}\del_{\mu}, &
\quad \{D_{\alpha},\Db_{\beta}\}&=-2i\sigma^{\mu}_{\alpha\beta}\del_{\mu},
\end{align}
with all the other anti-commutators vanishing. 
The supersymmetry transformation of a
superfield $\Phi(x,\theta,\thetab)$ is expressed as
\begin{align}
\label{n2sspace3}
 \delta \Phi(x,\theta,\thetab)&=(\xi Q-\overline{\xi} \Qb)\Phi.
\end{align}

\subsection{Supermultiplet}
\label{appsupermultiplet}
\subsubsection{Chiral multiplet}
Chiral superfield $\Phi(x,\theta,\thetab)$ is defined by the constraint
\begin{align}
\label{n2cm1}
 \Db_{\alpha}\Phi=0.
\end{align}
Using $y^{\mu}:=x^{\mu}+i\theta \sigma^{\mu} \thetab$, 
one can obtain the
component field representations:
\begin{align}
\label{n2cm2}
\Phi&=\Phi(y,\theta) \nonumber \\
&=\phi(y)+\sqrt{2}\theta\psi(y)+\theta\theta F(y) \nonumber \\
&=\phi(x)+i\theta\sigma^{\mu}\thetab \del_{\mu}\phi(x)-\frac14 \theta\theta \thetab\thetab \del^2 \phi(x)
+\sqrt{2}\theta\psi(x)+\frac{i}{\sqrt{2}}(\theta\theta) ( \thetab\sigma^{\mu}\del_{\mu}\psi(x))+\theta\theta F(x).
\end{align}
Similarly the anti-chiral superfield $\bar{\Phi}(x,\theta, \thetab)$ 
obeying the constraint $D_{m}\bar{\Phi}=0$ 
can be obtained from (\ref{n2cm2})
by conjugation:
\begin{align}
\label{n2cm3}
\bar{\Phi}
=\phib(x)-i\theta\sigma^{\mu}\thetab\partial_{\mu}\phib(x)
-\frac14 \theta\theta\thetab\thetab\partial^{2}\phib(x)
-\sqrt{2}\thetab\psib(x)
-\frac{i}{\sqrt{2}}(\thetab\thetab)(\theta\sigma^{\mu}\partial_{\mu}\psib(x))
-\thetab\thetab \bar{F}(x).
\end{align}

\subsubsection{Vector multiplet}
\label{appvm}
Vector superfield satisfies the relation
\begin{align}
\label{n2vm0}
V=\overline{V}.
\end{align}
Choosing the Wess-Zumino gauge we obtain
\begin{align}
\label{n2vm1}
V&=-\theta\sigma^{\mu}\bar{\theta}A_{\mu}
+i\theta\bar{\theta}\sigma
-i\theta\theta\thetab\bar{\lambda}
+i\thetab\thetab\theta\lambda-\frac12\theta\theta\thetab\thetab D(x).
\end{align}
One can express a field strength as a linear multiplet:
\begin{align}
\label{n2lmult1}
\Sigma:=-\frac{i}{2}\overline{D}DV.
\end{align}
In the components it is expressed as
\begin{align}
\label{n2lmult2}
\Sigma&=\sigma+\theta\bar{\lambda}-\lambda\thetab-i(\thetab\theta)D
+\frac{1}{2}(\thetab \epsilon \gamma^{\mu\nu}\theta)F_{\mu\nu} \nonumber \\
&-\frac{i}{2}\theta\theta (\thetab\sigma^{\mu}\partial_{\mu}\bar{\lambda})
+\frac{i}{2}\thetab\thetab(\theta\sigma^{\mu}\partial_{\mu}\lambda)+
\frac{1}{4}\theta\theta\thetab\thetab\partial^{\mu}\partial_{\mu}\sigma.
\end{align}

\bibliographystyle{utphys}
\bibliography{ref}

\end{document}